\documentclass[pdftex,twocolumn,epjc3]{svjour3}          

\RequirePackage[T1]{fontenc}

\smartqed  

\RequirePackage{graphicx}
\RequirePackage{mathptmx}      
\RequirePackage{flushend}
\RequirePackage[numbers,sort&compress]{natbib}
\RequirePackage[colorlinks,citecolor=blue,urlcolor=blue,linkcolor=blue]{hyperref}

\journalname{Eur. Phys. J. C}

\RequirePackage{widetext}
\RequirePackage{amsmath}
\RequirePackage{color}
\RequirePackage{float}
\RequirePackage{ulem}
\RequirePackage{amssymb}
\RequirePackage{amsfonts}
\RequirePackage{epsfig} 
\usepackage{amsmath} 
\RequirePackage{soul,xcolor}

\def\Rks{$R_{K^*}^{(s)}\ $}

\def\gbs{g_L^{bs}}
\def\gbd{g_L^{bd}}
\def\c9bsnp{$C_9^{bs,\rm NP}$}
\def\c10bsnp{$C_{10}^{bs,\rm NP}$}
\def\c9bdnp{$C_9^{bd,\rm NP}$}
\def\c10bdnp{$C_{10}^{bd,\rm NP}$}

\begin{document} 


\title{Predictions for $B_s \to \bar{K}^* \ell \,\ell$  in  non-universal  $Z'$ models }


\author{Ashutosh Kumar Alok\thanksref{e1,addr1}
  \and
Amol Dighe\thanksref{e2,addr2}
  \and
Shireen Gangal\thanksref{e3,addr2}
   \and
Dinesh Kumar \thanksref{e4,addr3,addr4}
}


\thankstext{e1}{e-mail: akalok@iitj.ac.in}
\thankstext{e2}{e-mail: amol@theory.tifr.res.in}
\thankstext{e3}{e-mail: shireen.gangal@theory.tifr.res.in}
\thankstext{e4}{e-mail: dinesh.kumar@ncbj.gov.pl}


\institute{Indian Institute of Technology Jodhpur, Jodhpur 342037, India \label{addr1}  
\and
Tata Institute of Fundamental Research, Mumbai 400005, India \label{addr2}  
 \and
National Centre for Nuclear Research,Pasteura 7, Warsaw 02-093, Poland \label{addr3} 
  \and
Department of Physics, University of Rajasthan, Jaipur 302004, India \label{addr4}
}

\date{Received: date / Accepted: date}

\maketitle

\begin{abstract}
The lepton flavor universality violating (LFUV) measurements $R_K$ and $R_{K^*}$ in $B$ meson decays can be accounted for in non-universal $Z'$ models. We constrain the couplings of these 
$Z'$ models by performing a global fit to correlated $b \to s \ell \ell$ and $b \to d \ell \ell $ processes, and calculate their possible implications for $B_s \to \bar{K}^*\ell \ell$ observables. For real new physics (NP) couplings,
the 1-$\sigma$ favored parameters allow the corresponding LFUV ratio $R_{K^*}^{(s)}$ in $B_s \to \bar{K}^*\ell \ell$ to range between 0.8 -- 1.2 at low $q^2$. Complex NP couplings improve the best fit only marginally, 
however they allow a significant enhancement of the branching ratio, while increasing the range of \Rks at low $q^2$ to 0.8 -- 1.8. We find that NP could cause zero-crossing in the forward-backward asymmetry $A_{FB}$
to shift towards lower $q^2$ values, and enhancement in the magnitude of integrated $A_{FB}$.  The $CP$ asymmetry $A_{CP}$ may be suppressed and even change sign. The simultaneous measurements of 
integrated \Rks and $A_{CP}$ values to 0.1 and 1\% respectively, would help in constraining the effective NP Wilson coefficient $C_9$ in $ b \to d \mu \mu$ interactions. 

\end{abstract}

\section{Introduction} 

In recent times, the most tenacious hints of physics beyond the standard model (SM) have been seen in the decays of $B$ mesons. In particular, there are several measurements in the decays
involving the quark-level transition $b\rightarrow s\, \ell^+\,\ell^-$ $(l=e,\,\mu)$ that deviate from the predictions of SM. These include the LFUV
observables $R_{K}$ and $R_{K^*}$  \cite{Hiller:2003js, Bordone:2016gaq} whose measurements disagree with the SM predictions at the level of $\sim 2.5\,\sigma$ \cite{rkstar,Rk2019}. This disagreement
can be attributed to NP in $b\rightarrow s\, e^+\,e^-$ and/or $b\rightarrow s\, \mu^+\,\mu^-$ \cite{Bhatia:2017tgo,Capdevila:2017bsm,Kumar:2019qbv}.
There are also deviations from the SM expectations at the level of $\sim 4\sigma$ in other measurements involving only $b \to s \mu^+ \mu^-$ transition,  such as the branching ratio of
$B_s \to \phi\, \mu^+\,\mu^-$ \cite{bsphilhc2} and angular observable $P'_5$ in $B \to K^* \, \mu^+\,\mu^-$ decay \cite{Kstarlhcb1,Kstarlhcb2, sm-angular}. Hence it is natural to try
accounting for the discrepancies in all the above measurements by assuming new physics only in the muon sector. 

These anomalies may be addressed in a model-agnostic way using the framework of effective field theory,  where the effects of NP are incorporated by adding new operators to 
the SM effective Hamiltonian. Various groups have performed global fits to all available data in the $b\to s \ell^+ \ell^-$ sector in order to identify the Lorentz structure of possible new physics operators \cite{Alguero:2019ptt,Alok:2019ufo,Ciuchini:2019usw,DAmico:2017mtc,Datta:2019zca,Aebischer:2019mlg,Kowalska:2019ley,Arbey:2019duh,Bhattacharya:2019dot,Coy:2019rfr}. 
Some of these new physics operators can be generated in $Z'$  \cite{Crivellin:2015lwa,Boucenna:2016wpr,Altmannshofer:2014cfa,Boucenna:2016qad,Darme:2018hqg} or leptoquark models \cite{Gripaios:2014tna,Fajfer:2015ycq,Varzielas:2015iva,Alonso:2015sja,
Calibbi:2015kma,Barbieri:2015yvd}.  It has been shown that several models with $Z'$, either light or heavy, can help account for the anomalies in $b\rightarrow s \mu^+ \mu^-$ sector \cite{{Chang:2010zy,Chang:2013hba,Buras:2013qja, Datta:2017pfz, Sierra:2015fma,
    Allanach:2015gkd,Calibbi:2019lvs}}.

Since the $Z'$ boson would in general couple to all generations, the imprints of such a $Z'$ would be seen in other flavor sectors as well. Therefore, it is worth extending this model to include other related decays. This will provide further insights into the NP flavor structure. In this work, we consider possible observable effects of $Z'$ models on decays induced by the quark-level transition $b\rightarrow d\, \mu^+\,\mu^-$.

The  $b \to d \mu^+\, \mu^-$ transition gives rise to inclusive semi-leptonic decays ${\bar B} \to X_d \,\mu^+ \, \mu^-$ as well as exclusive semi-leptonic decays such as
${\bar B} \to (\pi^0,\,\rho)\,\mu^+ \,\mu^-$, $B^+ \to \pi^+\, \mu^+\, \mu^-$, and $B_s \to \bar{K}^* \mu^+ \mu^-$. Till recently, the only observed decay mode among these
was $B^+ \to \pi^+\, \mu^+\, \mu^-$ \cite{LHCb:2012de,Aaij:2015nea}, however now LHCb has reported an evidence for the decay $B_s \to \bar{K}^* \mu^+ \mu^-$  with a 
 measured branching ratio of $(2.9 \pm 1.1) \times 10^{-8}$ \cite{Aaij:2018jhg}.  
For other decays, we only have an upper bound on their branching ratios \cite{Wei:2008nv,Lees:2013lvs}.

A large number of $b \to d \mu^+ \mu^-$ decays, at the level of thousands or tens of thousands, would be observed after the LHC upgrade. For example, about  17000 $B^+ \to \pi^+\, \mu^+\, \mu^-$ events
are expected to be observed after collection of the full 300 $\rm fb^{-1}$ dataset.
For $B_s \to \bar{K}^* \mu^+ \mu^-$ decays, the full angular analysis is expected to be possible after the LHCb Upgrade-II dataset, where around 4300 events could be observed \cite{Cerri:2018ypt}.
This would enable the measurements of angular observables in $B_s \to \bar{K}^* \mu^+ \mu^-$ decays with a precision even better than the existing measurements of angular distributions
in $B_d \to {K}^* \mu^+ \mu^-$ decay.

Currently, as there are not many measurements in the $b \to d$ sector, a model-independent analysis would not be very useful in constraining new physics. However, in the context of specific models 
(like $Z'$), some of the couplings can be constrained from the $b \to s$ sector and neutrino trident production.
Therefore we choose this approach to constrain the effective couplings in the $b\rightarrow d\, \mu^+\, \mu^-$ sector, and identify potential observables in the $B_s \to \bar{K}^* \mu^+\, \mu^-$ decay where large new physics effects are possible.

The paper is organized as follows. In section~\ref{sec:II}, we introduce the $Z'$ model considered in this work and indicate how it can be constrained by available measurements. We then describe our 
fit methodology in Section~\ref{sec:III}. The fit results along with predictions of various $B_s \to \bar{K}^*  \mu^+\, \mu^-$ observables are presented in Section~\ref{sec:IV}. 
We summarize in Section~\ref{sec:V}. 
 
 \section{The $Z'$ model and sources of constraints}
\label{sec:II}
 
 In the non-universal $Z'$ model that we consider, the $Z'$ boson is associated with a new $U(1)'$ symmetry. 
 It couples to both left-handed and right-handed muons but not to leptons of other generations. 
It couples to both left-handed and right-handed quarks, however we assume its couplings to right-handed quarks to be flavor-diagonal,
thereby avoiding contribution of new chirality flipped operators to flavor changing neutral current (FCNC) decays \cite{Barger:2009eq, Barger:2009qs}.
The change in the Lagrangian density due to the addition of such a heavy $Z'$ boson is 
 \begin{equation}
 \Delta \mathcal{L}_{Z'} = J^{\alpha}Z'_{\alpha}\;,
 \end{equation}
where
\begin{eqnarray}
  J^{\alpha} &\supset & g^{\mu\mu}_L\, \bar{L}\gamma^{\alpha}P_L L + g^{\mu\mu}_R\, \bar{L}\gamma^{\alpha}P_R\, L  + g^{bd}_L\, \bar{Q}_1 \gamma^{\alpha}P_L Q_3 
  \nonumber\\
  &&
+ g^{bs}_L\, \bar{Q}_2 \gamma^{\alpha}P_L Q_3
  + h.c. \,.
\label{eq:Jalpha}
\end{eqnarray}
The right-hand side in eq.~(\ref{eq:Jalpha}) includes only the terms contributing to FCNC processes. Here $P_{L(R)} = (1\mp \gamma_5)/2$, $Q_{i}$ is
the $i^{th}$ generation of quark doublet, and $L = (\nu_{\mu}, \mu)^T$ is the second generation doublet. 
Further,  $g_{L(R)}^{\mu\mu}$ are the left-handed (right-handed) couplings of the $Z'$ boson to muons, and $g_L^{bq}$ to quarks. 
One can integrate out the heavy $Z'$ and get the relevant terms in the effective four-fermion Hamiltonian as,
\begin{widetext}
\begin{eqnarray}
  \mathcal{H}_{\rm eff}^{Z'} = \frac{1}{2M^2_{Z'}}J_{\alpha}J^{\alpha}& \supset & \frac{g^{bs}_L}{M^2_{Z'}} \left(\bar{s}\gamma^{\alpha}P_L b\right)
  \left[\bar{\mu}\gamma_{\alpha}\left(g^{\mu\mu}_L P_L 
   + g^{\mu\mu}_R P_R\right)\mu \right]  
   +  \frac{\left(g^{bs}_L\right)^2}{2M^2_{Z'}}\left(\bar{s}\gamma^{\alpha}P_L b\right)\left(\bar{s}\gamma_{\alpha}P_L b\right)
  \nonumber\\
  & & + \frac{g^{bd}_L}{M^2_{Z'}} \left(\bar{d}\gamma^{\alpha}P_L b\right)\left[\bar{\mu}\gamma_{\alpha}\left(g^{\mu\mu}_L P_L + g^{\mu\mu}_R P_R\right)\mu \right]
   + \frac{\left(g^{bd}_L\right)^2}{2M^2_{Z'}}\left(\bar{d}\gamma^{\alpha}P_L b\right)\left(\bar{d}\gamma_{\alpha}P_L b\right)
  \nonumber \\
 & & + \frac{g^{\mu\mu}_L}{M^2_{Z'}} \left(\bar{\nu}_{\mu}\gamma_{\alpha}P_L\nu_{\mu}\right) \left[\bar{\mu}\gamma^{\alpha}\left(g^{\mu\mu}_L P_L + g^{\mu\mu}_R P_R\right)\mu\right],
 \label{Leff}
\end{eqnarray}
\end{widetext}
where we have taken the down-type quarks in the quark-doublets $Q_i$ to be in the mass-flavor diagonal basis.
In eq.~(\ref{Leff}), the first (third) term corresponds to 
$b\rightarrow s (d) \mu^+\mu^-$ transitions, the second (fourth) terms give rise to 
$B_{s}$--$\bar{B}_{s}$ ($B_{d}$--$\bar{B}_{d}$) mixing,
whereas the fifth term contributes to the neutrino trident production $\nu_{\mu} N \rightarrow \nu_{\mu} N \mu^+\mu^-$ ($N$ = nucleus). 
Consequently, 
the products $g_L^{bs} g_{L,R}^{\mu\mu} $ ($g_L^{bd} g_{L,R}^{\mu\mu} $) are constrained by the $b \to s(d) \mu^+ \mu^-$ data, and individual
magnitudes $|g_{L}^{bs}|$ ($|g^{bd}_L|$) 
from the $B_{s}$--$\bar{B}_{s}$  ($B_{d}$--$\bar{B}_{d}$) mixing.  
The neutrino trident production puts limits on the individual muon couplings $g_{L,R}^{\mu\mu}$. 
We now discuss constraints on the $Z'$ couplings arising from each of the above measurements. 

\subsection{$b\rightarrow s \,(d) \mu^+\mu^-$  decays}

The effective Hamiltonian for  $b\rightarrow q \mu^+\mu^-$ transition in the SM is
\begin{align} \nonumber
  \mathcal{H}_{\rm eff}^{\rm SM} &= -\frac{ 4 G_F}{\sqrt{2}} V_{tq}^* V_{tb}
  \bigg[ \sum_{i=1}^{6}C_i \mathcal{O}_i\\
  &  + C^{bq}_7\frac{e}{16 \pi^2}[\overline{q} \sigma_{\mu \nu}
      (m_q P_L + m_b P_R)b] F^{\mu \nu}  + C^{bq}_8 {\mathcal O}_8 \nonumber\\
     & + C^{bq,\rm SM}_9 \frac{\alpha_{\rm em}}{4 \pi}
    (\overline{q} \gamma^{\mu} P_L b)(\overline{\mu} \gamma_{\mu} \mu)  \nonumber\\
    &
    + C^{bq,\rm SM}_{10} \frac{\alpha_{\rm em}}{4 \pi}
    (\overline{q} \gamma^{\mu} P_L b)(\overline{\mu} \gamma_{\mu} \gamma_{5} \mu)
    \bigg] \;,
\end{align}
where $G_F$ is the Fermi constant and $V_{ij}$ are the Cabibbo-Kobayashi-Maskawa
(CKM) matrix elements.
The Wilson coefficients (WC) $C_i$ of the four-fermi operators ${\cal O}_i$
encode the short-distance contributions to the Hamiltonian in the SM,
where the scale-dependence is implicit, i.e. $C_i \equiv C_i(\mu)$
and ${\cal O}_i \equiv {\cal O}_i(\mu)$. 
The operators ${\cal O}_i$ ($i=1,...,6,8$) contribute to these processes
through the modifications  $C_{7,9}(\mu)$ $\rightarrow$ \\
$ C_{7,9}^{\mathrm{eff}}(\mu,q^2)$, where $q^2$ is the invariant mass-squared of the final state muon pair.
We drop the superscript ``eff" from here on for the sake of brevity.  
Addition of the new $Z'$ boson to the SM particle spectrum modifies the WCs as $C^{bq}_{9,10} \rightarrow C^{bq,\rm SM}_{9,10} + C^{bq,\rm NP}_{9,10}$,
where
 \begin{eqnarray}
  C^{bq,\rm NP}_9 &=& -\frac{\pi}{\sqrt{2}G_F\alpha V_{tb}V^*_{tq}} \frac{g_L^{bq}(g_L^{\mu\mu}+g_R^{\mu\mu})}{M^2_{Z'}}\,, \nonumber\\
  C^{bq,\rm NP}_{10} &=& \frac{\pi}{\sqrt{2}G_F\alpha V_{tb}V^*_{tq}} \frac{g_L^{bq}(g_L^{\mu\mu}- g_R^{\mu\mu})}{M^2_{Z'}}\,.
  \label{bqllNP}
\end{eqnarray}
In the $Z'$ models, $C^{bs,\rm NP}_9$ and $C^{bs,\rm NP}_{10}$ are in general independent. Two of the one-parameter scenarios,  $C^{bs,\rm NP}_{10} =0$ (popularly known 
as  $C^{bs,\rm NP}_{9} <0$) and
 $C^{bs,\rm NP}_9  = - C^{bs, \rm NP}_{10} $, can be realized by substituting $g^{\mu\mu}_L = g^{\mu\mu}_R$ and \ $g^{\mu\mu}_R = 0$, respectively. 

\subsection{$B_{s(d)}-\bar{B}_{s(d)}$ mixing}

The dominant contribution to $B_q-\bar{B}_q$ mixing within the SM comes from the virtual top quark in the box diagram.
The $Z'$ boson contributes to $B_q-\bar{B}_q$ mixing at the tree-level. The combined contribution to $M^q_{12}$, the dispersive part of the box diagrams responsible
for the mixing, is 
\begin{equation}
M^q_{12}= \frac{1}{3}M_{B_q}f_{B_q}^2 \widehat{B}_{B_q}\left[N\,C_{\rm VLL}^{\rm SM} + \frac{\left(g^{bq}_L\right)^2}{2M_{Z'}^2}\right],
\end{equation}
where
\begin{eqnarray}
N &=& \frac{G_F^2 M_W^2}{16 \pi^2} \left(V_{tb}V^*_{tq}\right)^2\,,\nonumber\\
C_{\rm VLL}^{\rm SM} &=& \eta_B x_t \left[1+\frac{9}{1-x_t}-\frac{6}{(1-x_t)^2}-\frac{6x_t^2 \ln x_t}{(1-x_t)^3}\right],
\end{eqnarray}
with $x_t \equiv m_t^2/M_W^2$. Here $\eta_B$=0.84 is the short-distance QCD correction calcualated at NNLO \cite{Buchalla:1995vs}, $f_{B_q}$ is the decay constant, and
$\widehat{B}_{B_q}$ is the bag factor.
The mass difference $\Delta M_q = 2|M_{12}^q|$ is
\begin{equation}
\Delta M_q = \Delta M_q^{\rm SM} \left|1+\frac{\left(g^{bq}_L\right)^2}{2\, N \,C_{\rm VLL}^{\rm SM} \,M_{Z'}^2}\right|\,,
\label{eq: Bdmix}
\end{equation}
while the relevant weak phase $\phi_q$ is
\begin{equation}
\phi_q = -2 \beta_q= {\rm arg}(M^q_{12})\,.
\end{equation}

\subsection{Neutrino trident production}

Within the $Z'$ models, the modification of the cross section $\sigma$ for neutrino trident production, $\nu_{\mu} N \rightarrow \nu_{\mu} N \mu^+\mu^-$ may be parameterized as
\cite{Alok:2017jgr}
\begin{eqnarray}
R_\nu =  \frac{\sigma}{\sigma_{\rm SM}} &=& \frac{1}{1+(1+4s^2_W)^2}\Bigg[\left(1+ \frac{v^2g^{\mu\mu}_L(g^{\mu\mu}_L-g^{\mu\mu}_R)}{M^2_{Z'}}\right)^2 \nonumber\\
&&
+ \left(1+4s^2_W+\frac{v^2g^{\mu\mu}_L(g^{\mu\mu}_L+g^{\mu\mu}_R)}{M^2_{Z'}}\right)^2\Bigg],
\label{trident}
\end{eqnarray}
where $v=246$ GeV and $s_W = \sin\,\theta_W$.

 \section{Fit Methodology}
\label{sec:III}
 
 We now determine favored values of the new physics couplings $g^{bs}_L$, $g^{bd}_L$, $g^{\mu\mu}_L$ and $g^{\mu\mu}_R$. 
 We nominaly take the mass of the $Z'$ boson to be $M_{Z'} = 1 \ \mathrm{TeV}$.  Note that since $M_{Z'}$
only appears through the combination $g^2/M_{Z'}^2$, the constraints on couplings can be appropriately scaled with the actual value of $M_{Z'}$. 

In $b\rightarrow s\mu^+\mu^-$ decays, we consider the following observables:
(i) $B_s \to \mu^+ \mu^-$ branching ratio \cite{Aaij:2013aka,CMS:2014xfa,Aaboud:2018mst}, 
(ii)   the updated value of $R_K$ by the LHCb collaboration~\cite{Rk2019}, 
(iii) $R_{K^*}$ measured by LHCb~\cite{rkstar} and its new Belle measurements, reported at Moriond'$19$~\cite{rkstar2019} (for Belle results, we use measurements
  in the bins $0.045$ ${\rm GeV}^{2}$ $<$ $q^2$ $<$ $1.1$ ${\rm GeV}^2$, $1.1 \,{\rm GeV}^{2}$ $<$ $q^2$ $<$ $6.0\, {\rm GeV}^2$,  and $15.0$ ${\rm GeV}^{2}$ $<$ $q^2$ $<$ $19.0$ ${\rm GeV}^2$, for $B^0$ as well
  as $B^+$ decays),
(iv)  the differential branching ratios of $B_d \to K^{*} \mu^+ \mu^-$~\cite{Aaij:2016flj,CDFupdate,Chatrchyan:2013cda,Khachatryan:2015isa},
$B^{+} \to K^{*+}\mu^{+}\mu^{-}$, $B_{d}\to K \mu^{+}\mu^{-}$,
$B^{+}\rightarrow K^{+}\mu^{+}\mu^{-}$ \cite{Aaij:2014pli,CDFupdate},
and $B \to X_{s}\mu^{+}\mu^{-}$ \cite{Lees:2013nxa}
in several $q^2$ bins,
(v) various $CP$-conserving and $CP$-violating angular observables in $B_d \to K^{*} \mu^+ \mu^-$
 \cite{kstaratlas,kstarcms,Khachatryan:2015isa,Kstarlhcb2,CDFupdate},
(vi) the measurements of differential branching ratio and angular observables of $B_{s}\to \phi \mu^{+}\mu^{-}$ \cite{bsphilhc2} in several $q^2$ bins.

While the ratios $R_K$ and $R_{K^*}$ are theoretically clean, other observables are plagued by sizeable uncertainties mainly coming from form factors.
  For $B_s \to \phi$ and $B \to K$ decays, we use the
  most precise form factor predictions obtained in light cone sum rule (LCSR)~\cite{Straub:2015ica, Gubernari:2018wyi}, taking into account the correlations between the uncertainties of different
  form factors and at different $q^2$ values. 
  The non-factorizable corrections are taken into account following the parameterization used in Ref.~\cite{Straub:2015ica,Straub:2018kue}. These are also
  compatible with the computations in Ref.~\cite{ Khodjamirian:2010vf}.

All the observables in the $b \to s \mu^+ \mu^-$ sector put constraints on the combinations $g_L^{bs}g^{\mu\mu}_L$ and $g_L^{bs}g^{\mu\mu}_R$. For the fit related to 
$b \to s \mu^+ \mu^-$, we closely follow the methodology of Ref.~\cite{Alok:2019ufo}. The  $\chi^2$  function for all the $b \to s \mu^+ \mu^-$ observables listed 
above is calculated as
\begin{equation}
\chi^2_{b \to s \mu \mu}(C_i) = [\mathcal{O}_{\rm th}(C_i) -\mathcal{O}_{\rm exp}]^T  \mathcal{C}^{-1} 
[\mathcal{O}_{\rm th}(C_i) -\mathcal{O}_{\rm exp}],
\label{chi-2d}
\end{equation} 
where  $C_i = C^{bs,\rm NP}_{9,10}$. Here $\mathcal{O}_{\rm th}(C_i)$ are the theoretical predictions of $b \to s \mu^+ \mu^-$ observables calculated using {\tt flavio} \cite{Straub:2018kue}, and
$\mathcal{O}_{\rm exp}$ are the corresponding experimental measurements. 
The total covariance matrix $\mathcal{C}$ is obtained by adding the individual theoretical and experimental covariance matrices.
In order to get the theoretical uncertainties, including the correlations among them, all input parameters such as the form factors, bag parameters, masses of particles, decay constants etc. are varied assuming a gaussian distribution, following the same methodology as used in {\tt{flavio}}~\cite{Straub:2018kue}. For the experimental covariance, we take into account 
the correlations among the angular
observables in $B \to K^{(*)} \mu^+ \mu^-$ \cite{Kstarlhcb2}
and $B_s \to \phi \mu^+ \mu^-$ \cite{bsphilhc2}.
For the other observables, we add the statistical and systematic errors in quadrature. Wherever the errors are asymmetric, 
we use the conservative approach of using the larger error on both sides of the central value.


We now turn to $B_q - \bar{B}_q$ mixing. Here we consider constraints from $\Delta M_d$, $\Delta M_s$, and the two CP-violating phases. 
Using $f_{B_d} \sqrt{\widehat{B}_{B_d}}= (225 \pm 9)$ MeV~\cite{Aoki:2019cca}, 
along with other input parameters from ref.~\cite{pdg}, eq.~(\ref{eq: Bdmix}) gives $\Delta M_d^{\rm SM}= (0.547 \pm 0.046)\,{\rm ps^{-1}}$. 
With $\Delta M_d^{\rm exp}=(0.5065 \pm 0.0019)\,{\rm ps^{-1}}$ \cite{Amhis:2019ckw}, 
the contribution of $\Delta M_d$ to $\chi^2$ is
\begin{equation}
\chi^2_{\Delta M_d} = \left(\frac{\Delta M_d - \Delta M_d^{\rm exp, m}}{\sigma_{\Delta M_d}}\right)^2\,,
\end{equation}
where we denote the experimental mean value of an observable $X$ by $X^{\rm exp,m}$, and the uncertainty in the observable by $\sigma_X$. 
In order to obtain $\sigma_X$, we add the experimental and theoretical uncertainties in quadrature. 
Here, $\sigma_{\Delta M_d}$ is dominated by the theoretical uncertainty.

In order to minimize the impact of theoretical uncertainties, we use $\Delta M_s$ constraints through the ratio $M_R = \Delta M_d / \Delta M_s$. 
In the SM, 
\begin{equation}
M_R^{\rm SM} 
= \left|\frac{V_{td}}{V_{ts}}\right|^2 \frac{1}{\xi^2}\frac{M_{B_d}}{M_{B_s}}\,,
\end{equation}
where $\xi =  \frac{f_{B_d}^2{\widehat{B}_{B_d}}}{f_{B_s}^2{\widehat{B}_{B_s}}}$. Using $\xi  = 1.2014^{+0.0065}_{-0.0072}$~\cite{King:2019lal}
and $\left|V_{td}/V_{ts}\right|$ = $0.2088^{+0.0016}_{-0.0030}$~\cite{Charles:2004jd}, we obtain
$M_R^{\rm SM} = 0.0297 \pm 0.0009$,
where we have added the errors in quadrature. Wherever there are asymmetric errors, we take a conservative approach and use the larger of the errors on two sides.
The value of $M_R^{\rm exp} = 0.0285 \pm 0.0001$ \cite{Amhis:2019ckw}, so the contribution to $\chi^2$ due to this ratio is
\begin{equation}
\chi^2_{M_R} = \left(\frac{M_R - M_R^{\rm exp, m}}{\sigma_{ M_R}}\right)^2\,.
\end{equation}
The observables $\Delta M_d$ and $M_R$  constrain $|g_L^{bs}|$ and $|g_L^{bd}|$.

The $CP$-violating constraints from $J/\psi \phi$ and $J/\psi K_S$ decays contribute to the $\chi^2$ as
\begin{eqnarray}
\chi^2_{J/\psi\phi} &=& \left(\frac{S_{J/\psi\phi} - S_{J/\psi\phi}^{\rm exp,m}}{\sigma_{S_{J/\psi\phi}}} \right)^2\,,\nonumber\\
\chi^2_{J/\psi K_S} &=& \left(\frac{S_{J/\psi K_S} - S_{J/\psi K_S}^{\rm exp,m}}{\sigma_{S_{J/\psi K_S}}} \right)^2,
\end{eqnarray}
where $S_{J/\psi\phi}$ =  - Im$[M_{12}^s]/|M_{12}^s|$ and $S_{J/\psi K_S}$ =  Im$[M_{12}^d]/|M_{12}^d|$. Here we have taken the measurements to be $S_{J/\psi\phi}^{\rm exp} =0.02 \pm 0.03$ and $S_{J/\psi K_S}^{\rm exp} = 0.69 \pm 0.02$ \cite{pdg}.

For the constraints from neutrino trident production, we use the quantity $R_\nu = \sigma/\sigma_{\rm SM}$, whose theoretical expression  is given in eq.~(\ref{trident}).  We have taken $R_\nu^{\rm exp} \equiv  0.82\pm 0.28$~\cite{Mishra:1991bv, Altmannshofer:2019zhy}.
The contribution to the total $\chi^2$ is
\begin{equation}
\chi^2_{\rm {trident}} = \left(\frac{R_\nu - R_\nu^{\rm exp,m}}{\sigma_{R_\nu}}\right)^2\,.
\end{equation}
This observable constraints $g_L^{\mu \mu}$ and $g_R^{\mu \mu}$.

The $b \to d \mu^+\, \mu^-$ decays are CKM-suppressed as compared to $b \to s \mu^+\, \mu^-$. 
In our analysis, we include constraints from the branching ratios of  $B^+ \to \pi^+\, \mu^+\, \mu^-$ and
$B_d \to \mu^+\, \mu^-$ decays.  We do not include the measurements of observables in $B_s \to \bar{K}^* \mu^+ \mu^-$ decay in our fit, since we are interested in obtaining predictions for these.

The theoretical expression for ${\cal B}(B^+ \to \pi^+\, \mu^+  \mu^-)$ 
in the $Z'$ model can be obtained from Ref.~\cite{Wang:2007sp}, by adding the NP contribution as given in eq.~(\ref{bqllNP}). 
The contribution to $\chi^2$ from this decay is
\begin{equation}
\chi^2_{B^+ \to \pi \, \mu \, \mu }  =\Big( \frac{{\cal B}(B^+ \to \pi \mu \mu ) - {\cal B}(B^+ \to \pi \mu \, \mu)^{\rm exp,m}}
{\sigma_{{\cal B}(B^+ \to \pi \mu  \mu)}} \Big)^2\;,
\end{equation}
where $ {\cal B}(B^+ \to \pi  \mu \, \mu)^{\rm exp} = (1.83 \pm 0.24)\times 10^{-8}$ \cite{Aaij:2015nea}. Following Ref.~\cite{Wang:2007sp},
a theoretical error of $15\%$ is included due to uncertainties
in the $B \to \pi$ form factors \cite{Ball:2004ye}. 

The branching ratio of $B_d \to \mu^+ \,\mu^-$ in our model is given by
\begin{eqnarray}
{\cal B}(B_d \to \mu^+ \,\mu^-) &=& \frac{G^2_F \alpha^2 M_{B_d} m_\mu^2 f_{B_d}^2 \tau_{B_d}}{16 \pi^3} 
|V_{td}V^*_{tb}|^2  \nonumber\\
& \times &
 \sqrt{1 - \frac{4 m_\mu^2}{M_{B_d}^2}} \left|{C}_{10}^{bd,\rm SM} + C^{bd,NP}_{10} \right|^2,
\end{eqnarray}
and the contribution to $\chi^2$ is 
\begin{equation}
\chi^2_{B_d \to \, \mu \, \mu }  =\Big( \frac{{\cal B}(B_d \to \mu \mu) - {\cal B}(B_d \to \mu \mu)^{\rm exp,m} }
{\sigma_{{\cal B}(B_d \to \mu \mu)}} \Big)^2\;.
\end{equation}
We have used ${\cal B}(B_d \to \mu^+ \,\mu^-)^{\rm exp}= (3.9\pm 1.6)\times 10^{-10}$~\cite{Amhis:2019ckw}, $f_{B_d}=(190 \pm 1.3)$MeV \cite{King:2019lal}, and other inputs from \cite{pdg}.

 \begin{figure*}[t]
 \begin{center}
 \includegraphics[width=0.5\textwidth]{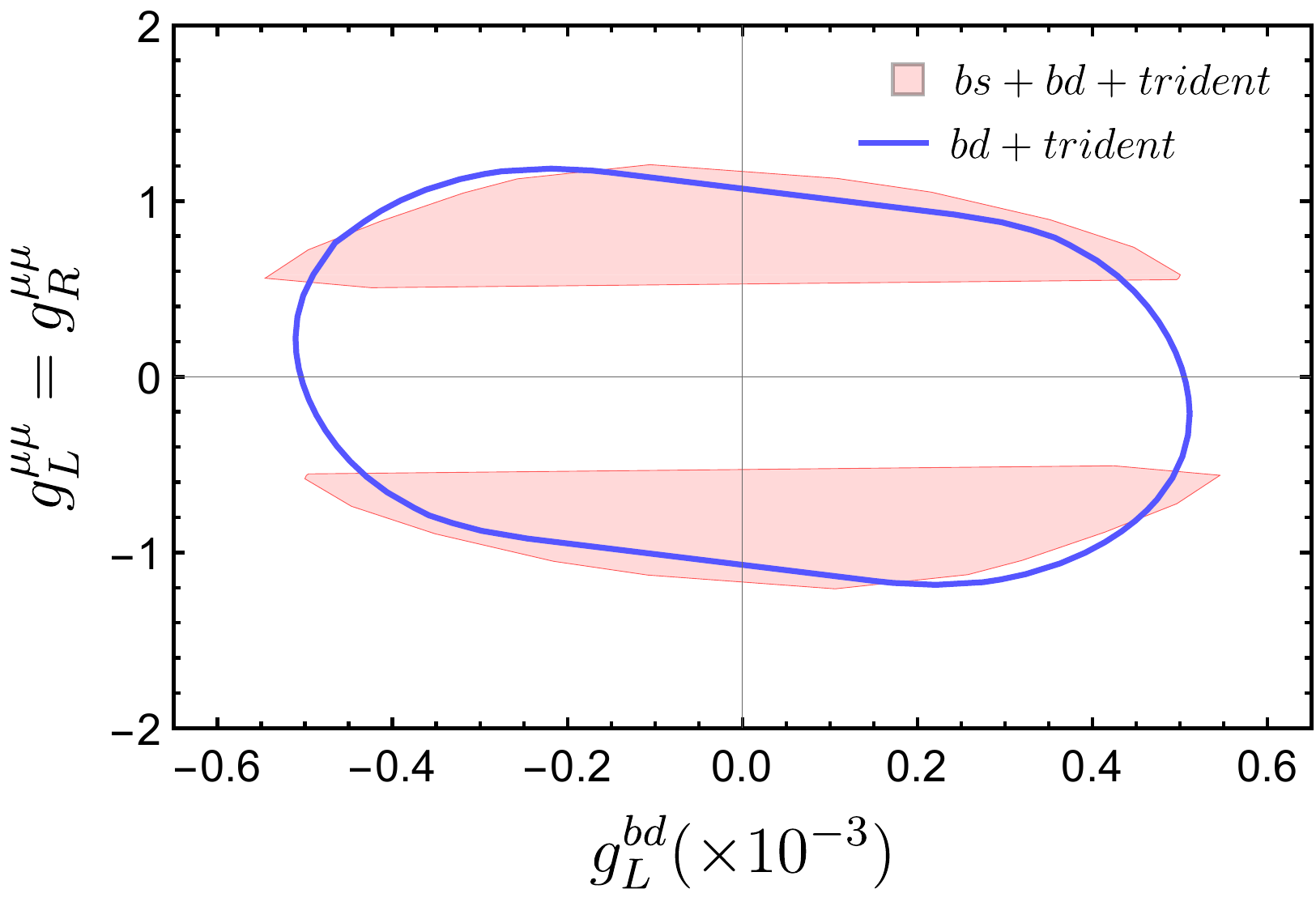}
\caption{The $(\gbd,\,g_L^{\mu\mu})$ parameter space corresponding to a $Z'$ model with $g_L^{\mu\mu} = g_R^{\mu\mu}$  (i.e. $C_{10}^{bd,NP} =0$), for $M_{Z'}=1$ TeV.
  The blue curve is the boundary of the 1$\sigma$-favored region due to constraints from measurments in $b \to d$ sector and neutrino trident production. 
  The pink shaded region represents the $1\sigma$-favored parameter space after including additional constraints from $b \to s \mu^+\, \mu^-$ data and $B_s - \bar{B_s}$ mixing.}
\label{compare}
 \end{center}
 \end{figure*}

Finally, combining all the above constraints, we obtain
\begin{eqnarray}
  \chi^2_{\rm total} &=& \chi^2_{b\to s \mu \mu} + \chi^2_{\Delta M_d} + \chi^2_{M_R} + \chi^2_{J/\Psi \phi} + \chi^2_{J/\Psi K_S} + \chi^2_{\rm trident}
 \nonumber\\
 &&+ \chi^2_{B^+ \to \pi \, \mu \, \mu} + \chi^2_{B_d \to \mu \mu}\,.
\end{eqnarray}

In addition to the above constraints, there would be constraints coming from $b \to s \nu \, \bar{\nu}$ and charm sector.  However, at present,  we only have upper limits on $b \to s\, \nu \, \bar{\nu}$ and $c \to u \, \mu^+ \,\mu^-$ decays. Further, in the $D^0$-$\bar{D^0}$ mixing, we expect a large fraction of unknown long-distance contributions. Therefore 
these measurements cannot be included in as clean a manner as the ones we have considered above. Instead, in the appendices A and B, we determine the allowed regions due to these constraints taken separately  and compare with those obtained from our fit.  We find that the constraints from these additional channels are much weaker, and will not affect our results.

In the next section, we present our fit results, along with predictions of several observables in $B_s \to \bar{K}^* \mu^+ \mu^-$ decay.

 \section{Fit results and predictions}
\label{sec:IV}

In a model-independent analysis, there have been attempts to put limits on the new physics couplings for $b \to d \ell \ell$ decays \cite{Rusov:2019ixr}, however it is
difficult as there are only a few measurements in this sector.  
 Within the context of a $Z'$ model, one can obtain meaningful constraints using correlated $b \to s$ and $b \to d$ processes.
 This can be seen from Fig.~\ref{compare}, which depicts the allowed $(\gbd,\,g_L^{\mu\mu}= g_R^{\mu \mu} )$ parameter space corresponding to a $Z'$ model which generates the 1D scenario
 $C_{10}^{bd,\rm NP} = 0$. The elliptical region represents the 1$\sigma$-favored parameter space with constraints only from $b \to d$ sector, i.e.,
 branching ratios of $B^+ \to \pi^+\, \mu^+\, \mu^-$ and
 $B_d \to \mu^+ \mu^-$ decays, $B_d - \bar{B_d}$ mixing, and neutrino trident production. The two shaded regions represent the 1$\sigma$-favored parameter space obtained by including additional constraints
 from all relevant measurements related to $b \to s \mu^+\, \mu^-$ decays and $B_s - \bar{B_s}$  mixing. It can be seen that the allowed range of NP couplings,
 in particular $g_{L,R}^{\mu \mu}$, reduces considerably after including constraints from the $b \to s$ sector.  Therefore, it is worth studying implications of several measurements in the $b \to s$ sector 
on the observables in $b \to d \mu^+\, \mu^-$ decays.

 Performing a fit to the relevant observables in $b \to s$ and $b \to d $ sectors, we determine the 1$\sigma$-favored parameter space of the couplings $g_L^{bd}$, $g_L^{bs}$, $g_L^{\mu \mu}$ and $g_R^{\mu \mu}$,
 considering $g_L^{bs}$ and $g_L^{bd}$ to be (i) real, (ii) complex. These can be used to 
 find constraints on the NP Wilson coefficients ($C_9^{bd,\rm NP},C_{10}^{bd, \rm NP}$),
 and to put limits on the allowed NP in the following observables in $B_s \to \bar{K}^* \mu^+ \mu^-$ decay: differential branching ratio, the LFUV ratio \Rks, muon forward-backward asymmetry $A_{FB}$,
 longitudinal polarization fraction $F_L$, and direct $CP$ asymmetry $A_{CP}$.

\begin{figure*}[t]
\includegraphics[width=0.4\textwidth]{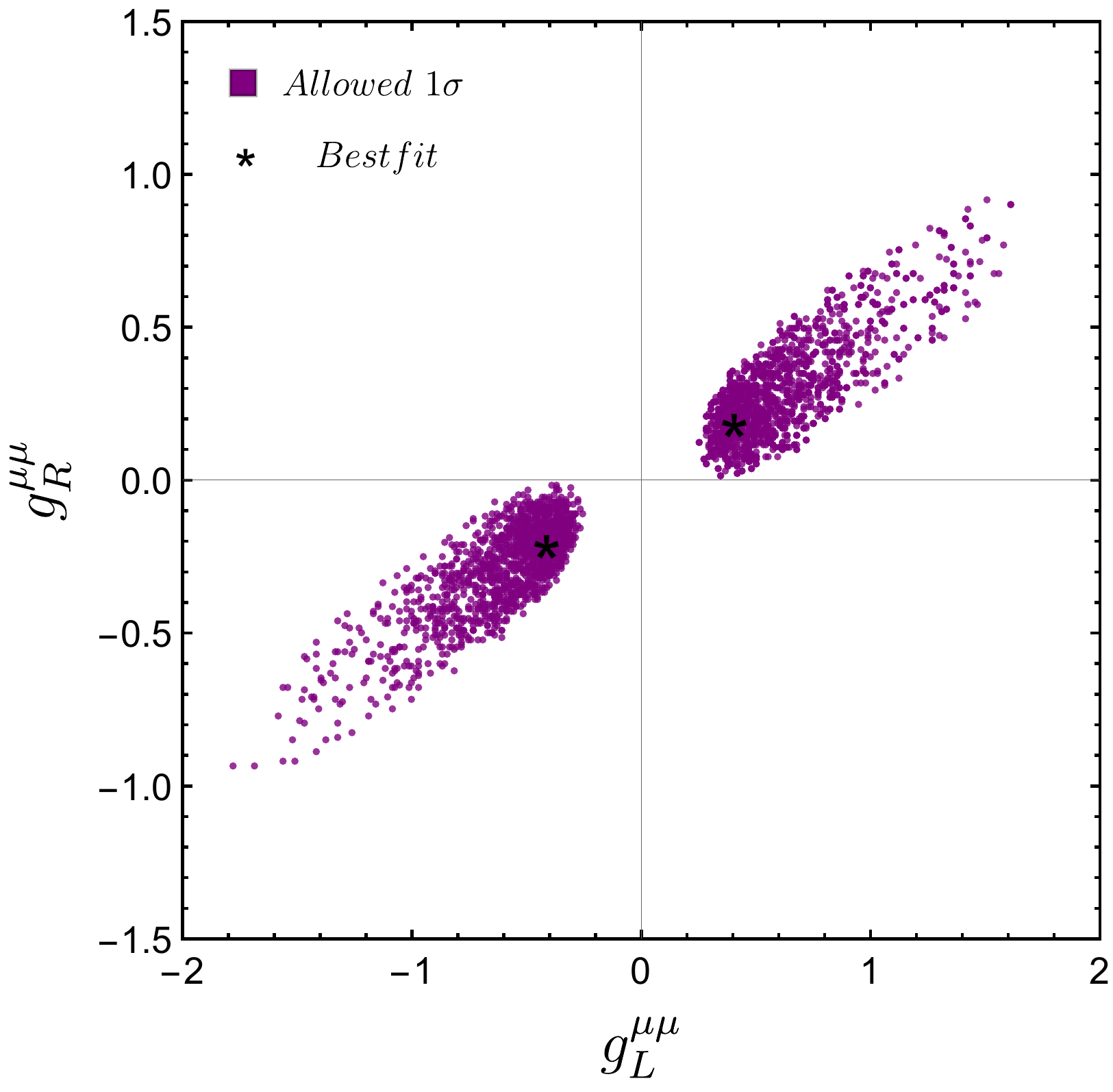}
\hspace{0.6cm}
\includegraphics[width=0.4\textwidth]{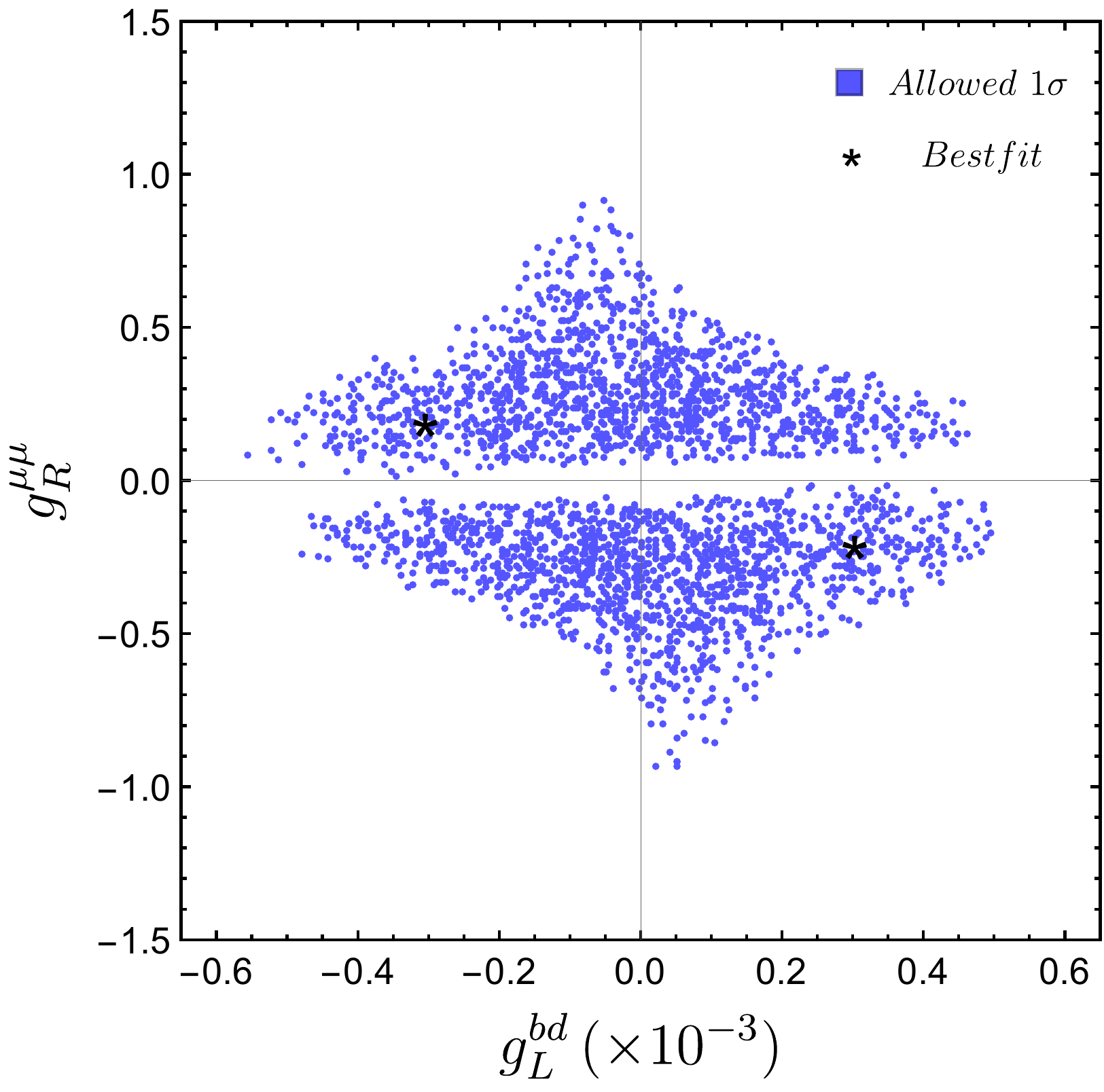}\\
\vspace{0.4cm}
\includegraphics[width=0.4\textwidth]{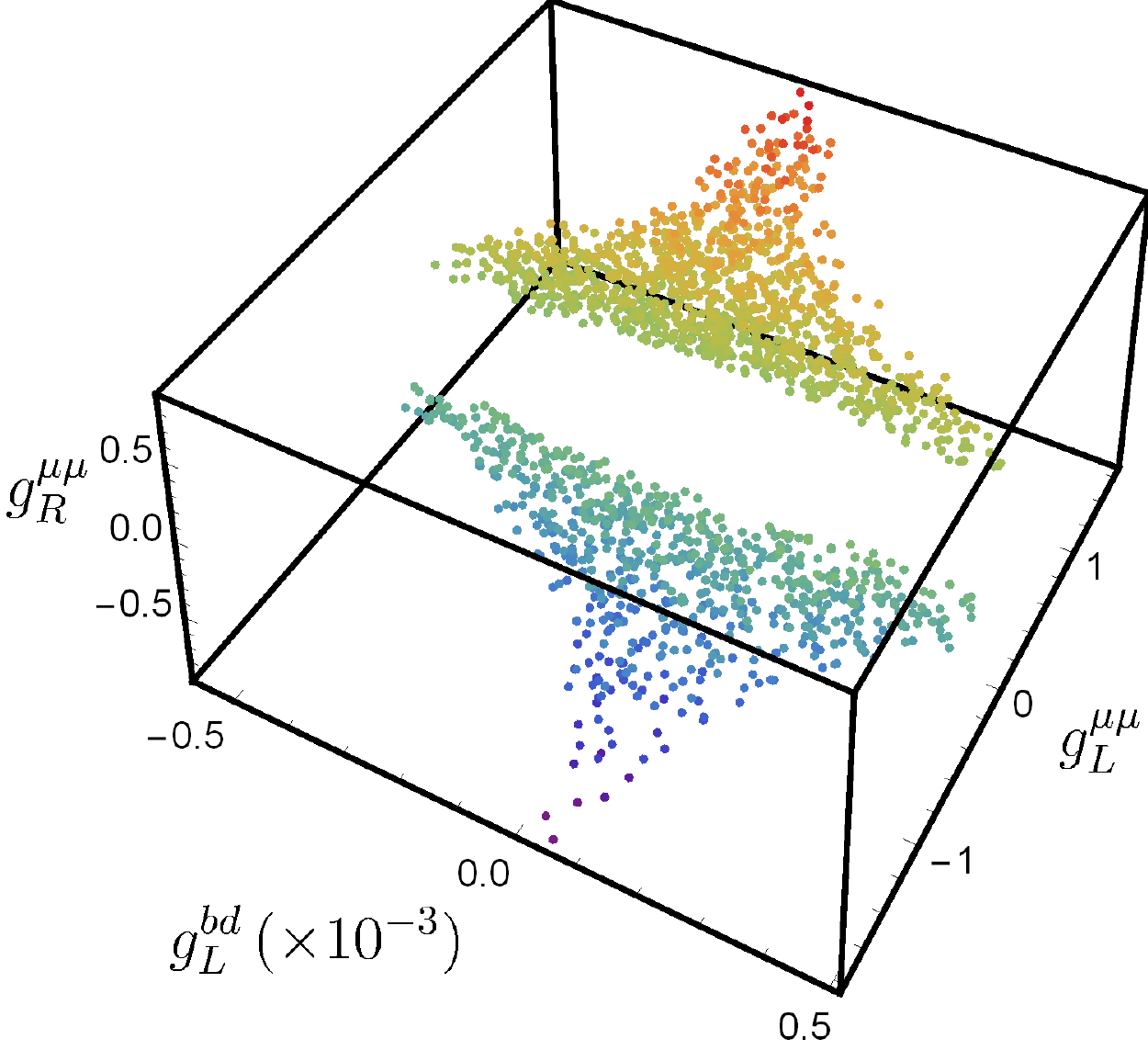}
\hspace{0.6cm}
\includegraphics[width=0.4\textwidth]{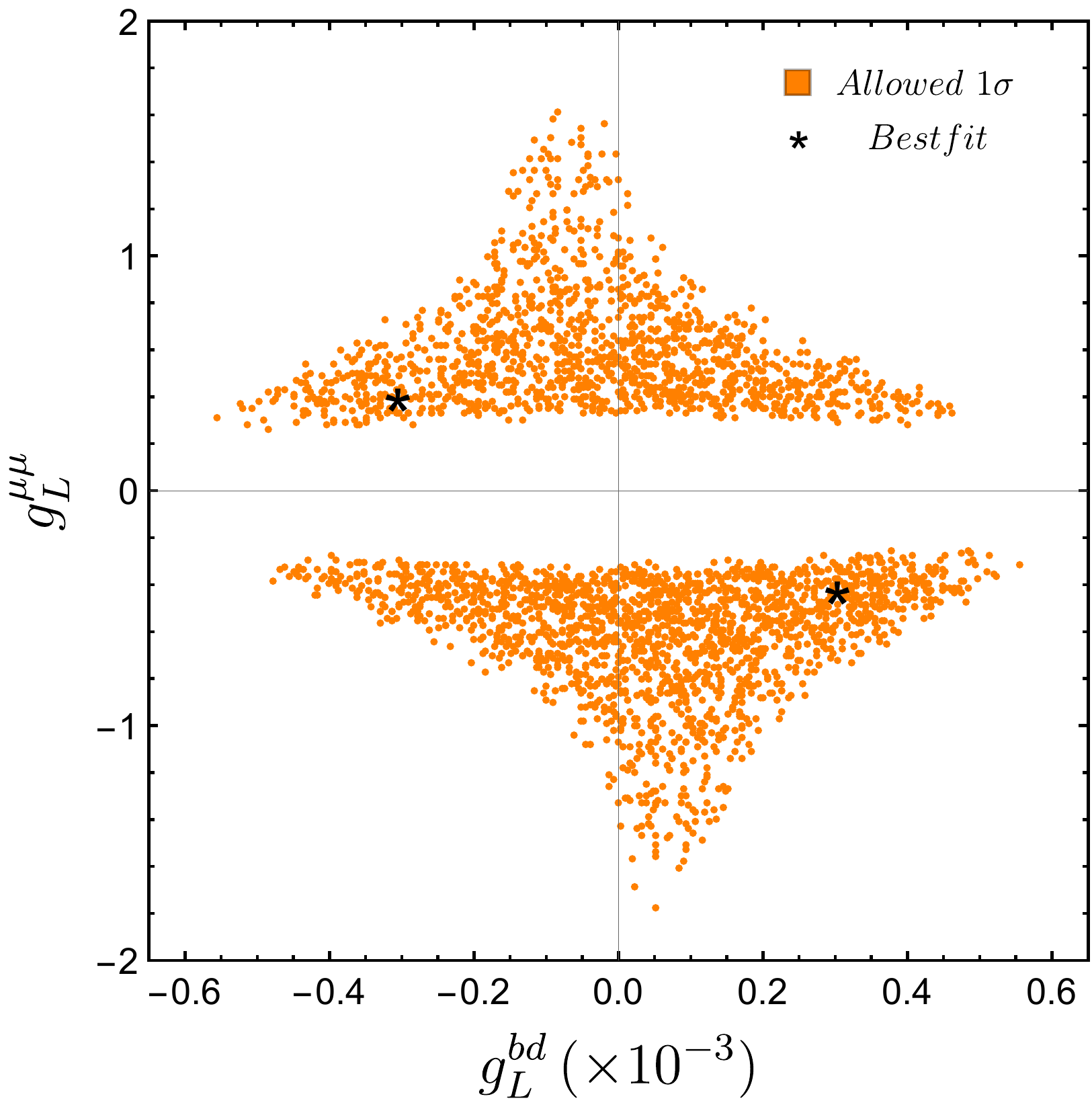}
\caption{The 1$\sigma$-favored ($g^{bd}_L $, $g_{L}^{\mu\mu}$, $g_{R}^{\mu\mu}$) parameter space for a $Z'$ model with real couplings, for $M_{Z'}=1$ TeV. The colors red to blue in the bottom
left 3D parameter space correspond to decreasing values of $g_R^{\mu \mu}$. }
\label{parameterS-real}
 \end{figure*}

The matrix element for the decay amplitude of $B_s \to \bar{K}^*\mu^+\mu^-$ can be written as
\begin{align}\label{eqnHamil}
  \mathcal{M}=&\frac{G_F \alpha}{\sqrt{2}\pi}V_{tb}V_{td}^* \Big{\{}\Big[C_9^{bd}\left<\bar{K}^*|\bar{d}\gamma^{\mu} P_L b|B_s\right> \nonumber\\
	&
    -\frac{2m_b}{q^2}C_{7}^{bd}\left<\bar{K}^*|\bar{d}\,i \sigma^{\mu\nu}q_{\nu} P_R\, b|B_s\right>\Big](\bar{\mu}\gamma_{\mu}\mu)\nonumber\\
	&+C_{10}^{bd} \left<\bar{K}^*|\bar{d}\gamma^{\mu} P_L b|B_s\right>(\bar{\mu}\gamma_{\mu}\gamma_5\mu)\Big{\}}\,,
\end{align}
where $C_9^{bd, \rm SM}$ and $C_{10}^{bd, \rm SM}$ are taken from Ref. \cite{Alok:2008dj} and Ref. \cite{Asatrian:2003vq} respectively.   
The matrix elements appearing in eq.~(\ref{eqnHamil}) have been calculated using form factors obtained by a combined fit to lattice calculations and QCD sum rules on the light cone
~\cite{Straub:2015ica}. 
We also include the non-factorizable corrections due to soft gluon emission and charmonium resonance, which have been computed for $B_d \to K^* \ell \ell$
\cite{Descotes-Genon:2015uva, Khodjamirian:2010vf}, and
parameterized as corrections to $C_9^{\rm SM}$.  These effects are assumed to be roughly the same for $B_s \to \bar{K}^*\ell \ell $ due to flavor symmetry \cite{Kindra:2018ayz}.

 \begin{table*}[h]
  \begin{center}
\begin{tabular}{|c|c|c|c|c|}
\hline
\quad\quad  Scenario \quad\quad  & \quad \quad\ NP1 \quad\quad &\quad\quad NP2 \quad\quad  & \quad\quad NP3 \quad \quad & \quad\quad NP4 \quad\quad \\
 \hline
 \quad\quad $C_9^{bd, \rm NP}$ \quad\quad  &  \quad\quad $+0.98$ \quad\quad & \quad \quad $-0.80$ \quad \quad & \quad \quad $-1.4 +  4.9\, i$ \quad \quad & \quad  \quad $-0.6 + 0.8\,i$ \quad \quad
 \\
\hline
\quad \quad $C_{10}^{bd, \rm NP}$ \quad \quad &\quad \quad $-0.17$ \quad \quad & \quad \quad $+0.19$ \quad \quad &\quad \quad $+0.7 - 2.3\,i$ \quad \quad &\quad \quad  $+0.2 -0.2\,i$ \quad \quad \\ 
\hline
    \end{tabular}
        \caption{Values of NP Wilson coefficients for benchmark scenarios NP1, NP2 corresponding to a $Z'$ model with real couplings, and scenarios NP3, NP4 with complex couplings. The first two benchmark scenarios NP1 and NP2 for real $Z'$ couplings, correspond to a maximum deviation from the SM predictions of the observables considered. 
The last two scenarios, for complex couplings, are the 1$\sigma$-favoured ones
  with a near-maximum value of Im[$C_9^{bd,\rm NP}$] and a near-minimum value of Re[$C_{10}^{bd, \rm NP}$].} 
 \label{table-imwc}
  \end{center}
\end{table*}

\begin{figure*}[t]
\includegraphics[width=0.45\textwidth]{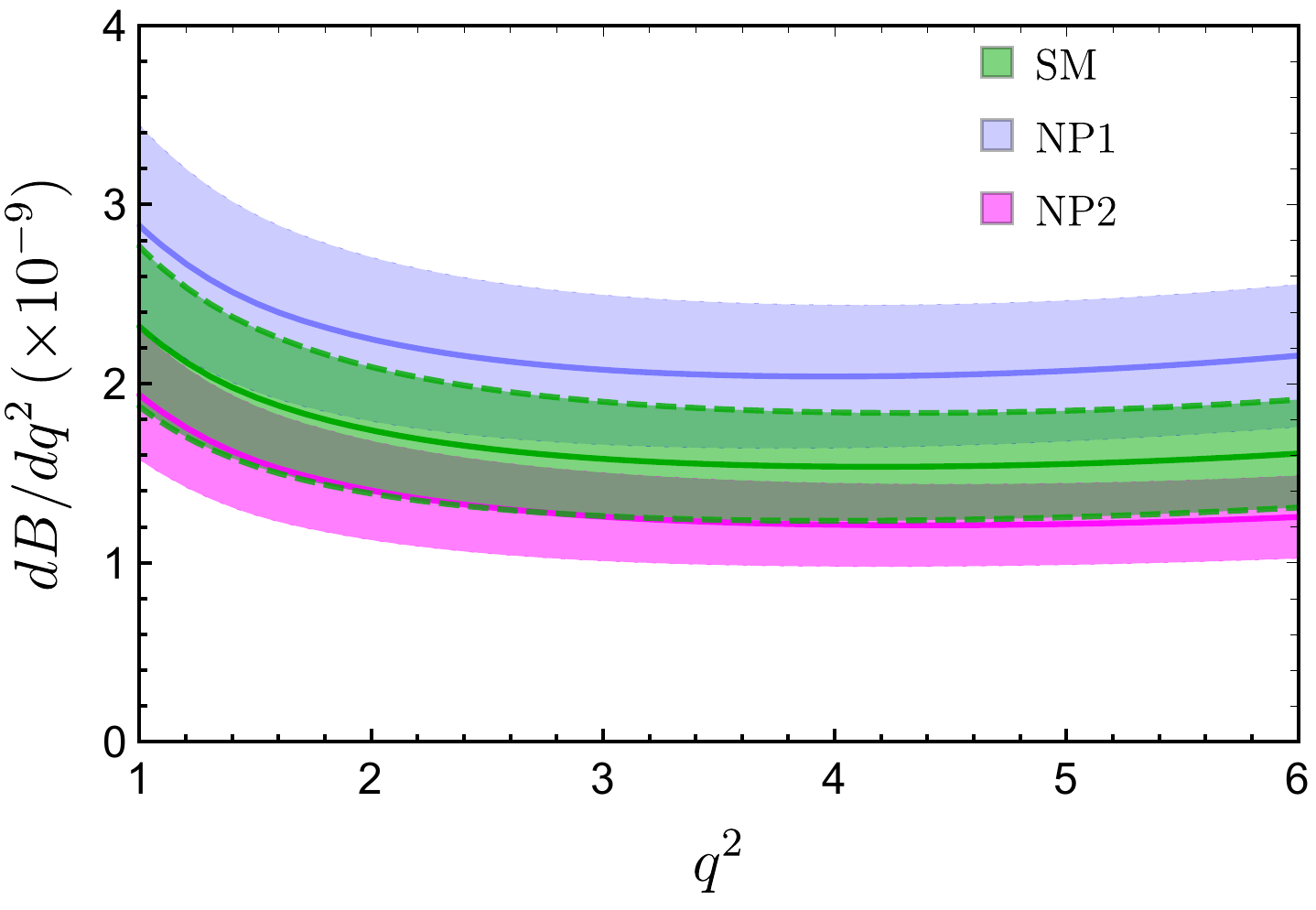}
\hspace{0.05\textwidth}
\includegraphics[width=0.455\textwidth]{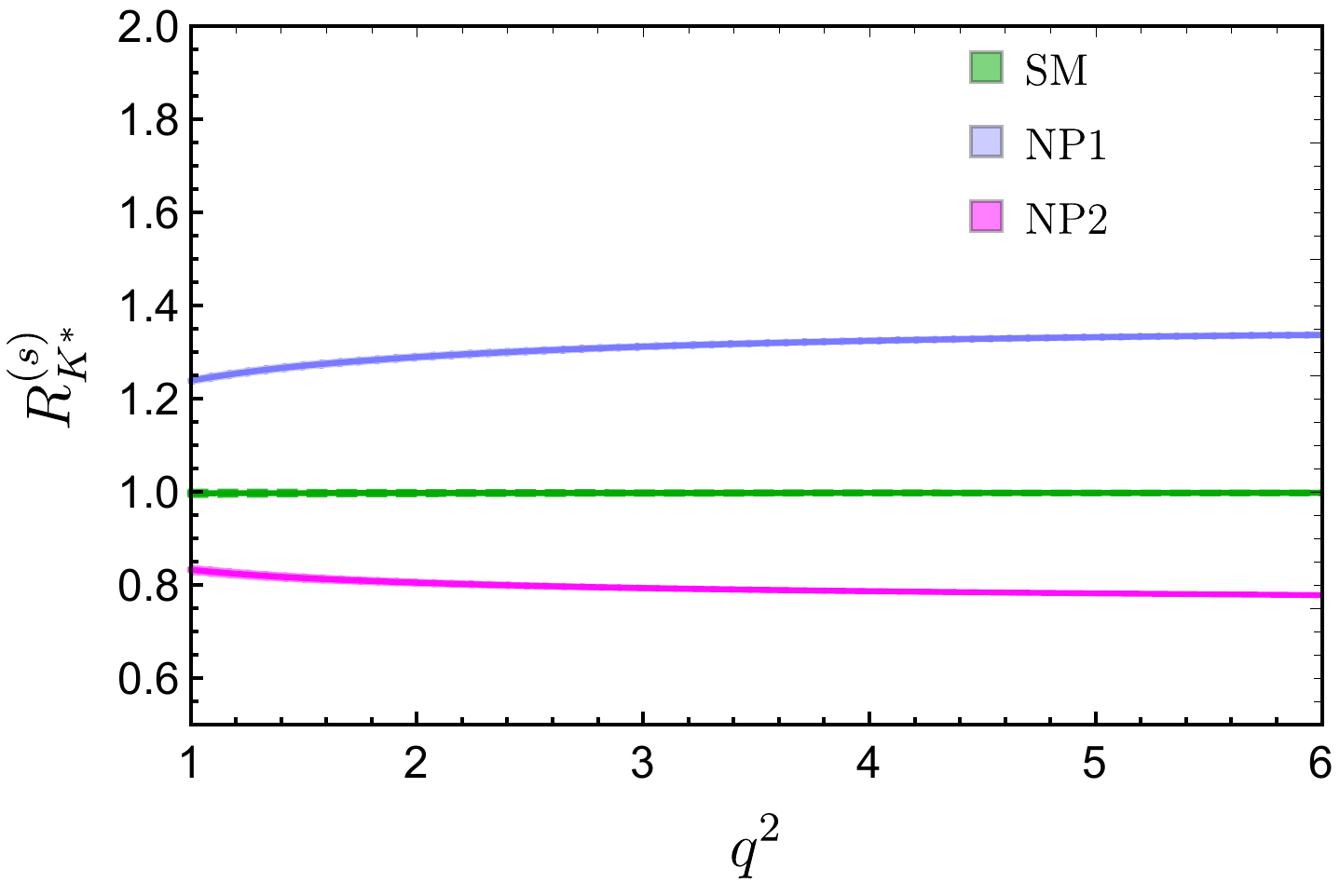}
\\
\includegraphics[width=0.46\textwidth]{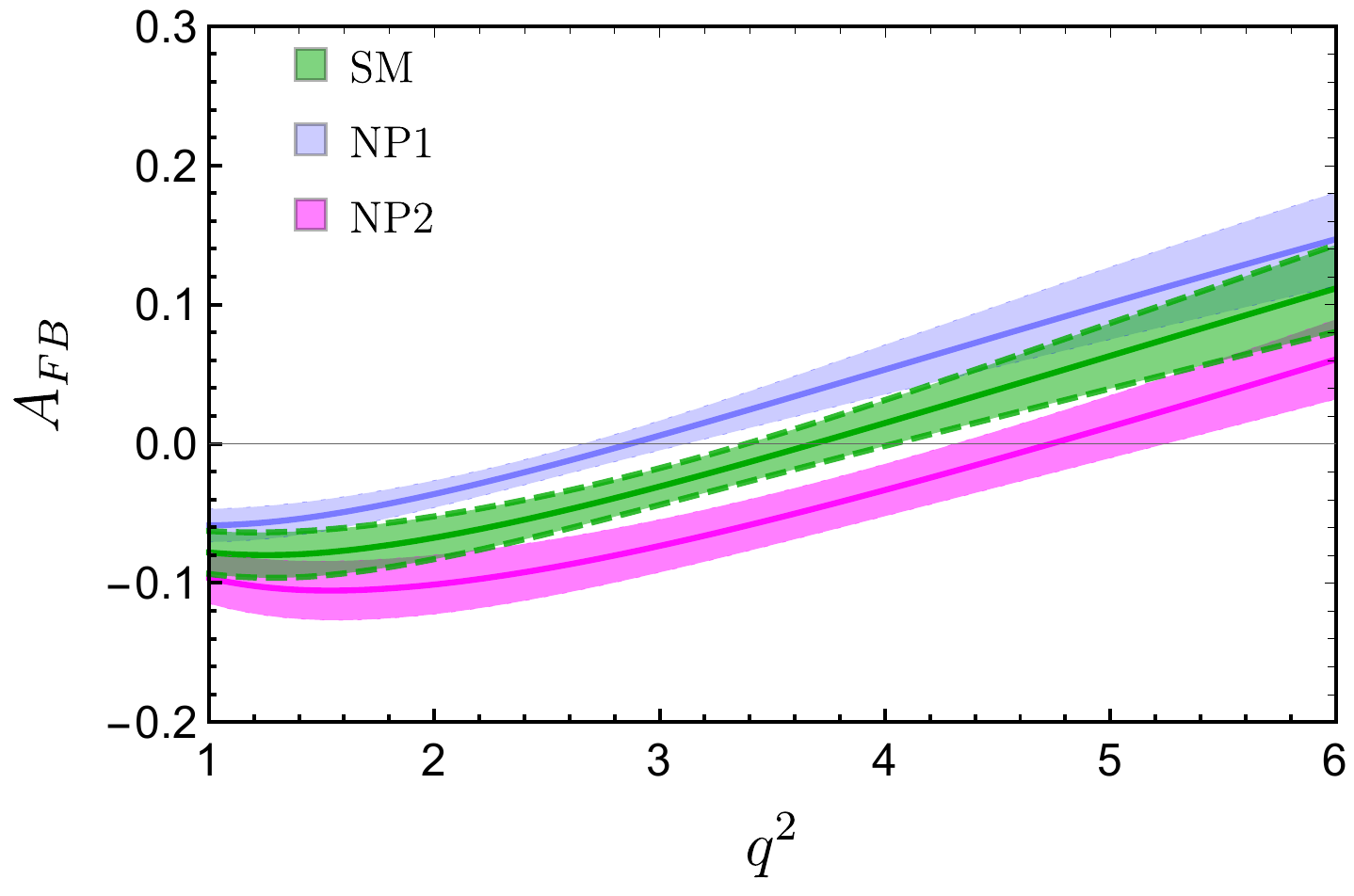}
\hspace{0.05\textwidth}
\includegraphics[width=0.45\textwidth]{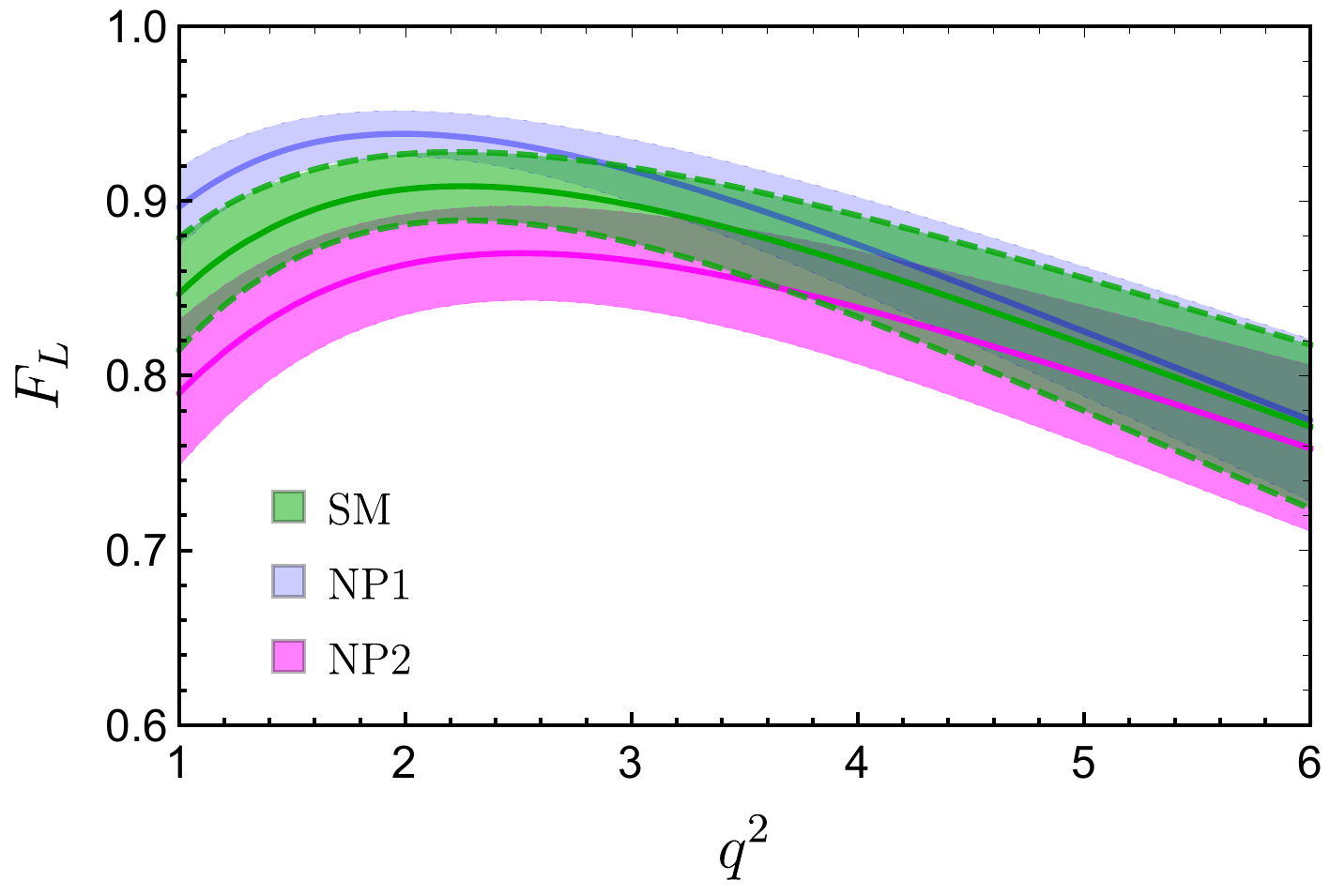}
\caption{The predictions for observables in $B_s \to \bar{K}^*\mu^+ \mu^-$ decay, with real $Z'$ couplings, for SM and the benchmark scenarios NP1 and NP2 as in Table~\ref{table-imwc}\,.}
\label{fig:re-pred}
\end{figure*}

The decay  $B_s \to \bar{K}^*\mu^+ \mu^-$ may be described in terms of the four-fold distribution as \cite{Kindra:2018ayz}
\begin{align}
  \frac{d\,\Gamma}{dq^2} &=  \int_{-1}^{+1} d\cos{\theta_l}\, d\cos{\theta_V}\, \int_0^{\pi} d\phi \frac{d^4\Gamma}{dq^2\, d\cos{\theta_V}\,d\cos{\theta_l}\,d\phi}\nonumber\\
 & =  \frac{1}{4} (3\,I_1^c + 6 I_1^s - I_2^c -2I_2^s)\,,
\end{align}
where $q^2$ is the lepton invariant mass, $\theta_V$ and $\theta_l$ are the polar angles, and $\phi$ is the angle between the dimuon plane and $K^*$ decay plane.
The relevant observables
can be obtained from the four-fold distribution as 
\begin{align}
   \frac{dB}{dq^2} &= \tau_{B_s} \frac{d\,\Gamma}{d\,q^2} \,, \nonumber\\ 
    R_{K^*}^{(s)}(q^2) &= \frac{d\,\Gamma(B_s \to \bar{K}^* \mu^+ \mu^- )/d\,q^2}{d\,\Gamma(B_s \to \bar{K}^* e^+ e^-)/d\,q^2}\,, \nonumber\\
   A_{FB}(q^2) &= \frac{1}{d\Gamma/dq^2}\Big[\int_{-1}^0-\int_0^1 \Big]d\cos{\theta_l}\frac{d^4 \Gamma}{dq^2 d\cos{\theta_l}} \nonumber\\&
    = \frac{-3 I_6^s}{3I_1^c + 6 I_1^s - I_2^c-2 I_2^s}\,, \nonumber \\
F_L(q^2) &= \frac{3I_1^c - I_2^c}{3I_1^c + 6I_1^s- I_2^c - 2I_2^s}\,, \nonumber\\ 
A_{CP}(q^2)&= \frac{dB/dq^2- d \bar{B}/dq^2}{dB/dq^2 + d\bar{B}/dq^2}\,,
\end{align}
where the functions $I_i$ can be expressed in terms of the transversity amplitudes ~\cite{Altmannshofer:2008dz}. Here $\bar{B}$ corresponds to the decay mode $\bar{B}_s \to K^* \mu^+ \mu^-$.

We present our results for the above observables at four benchmark NP scenarios as given in Table~\ref{table-imwc}. 

\begin{figure*}[t]
\includegraphics[width=0.46\textwidth]{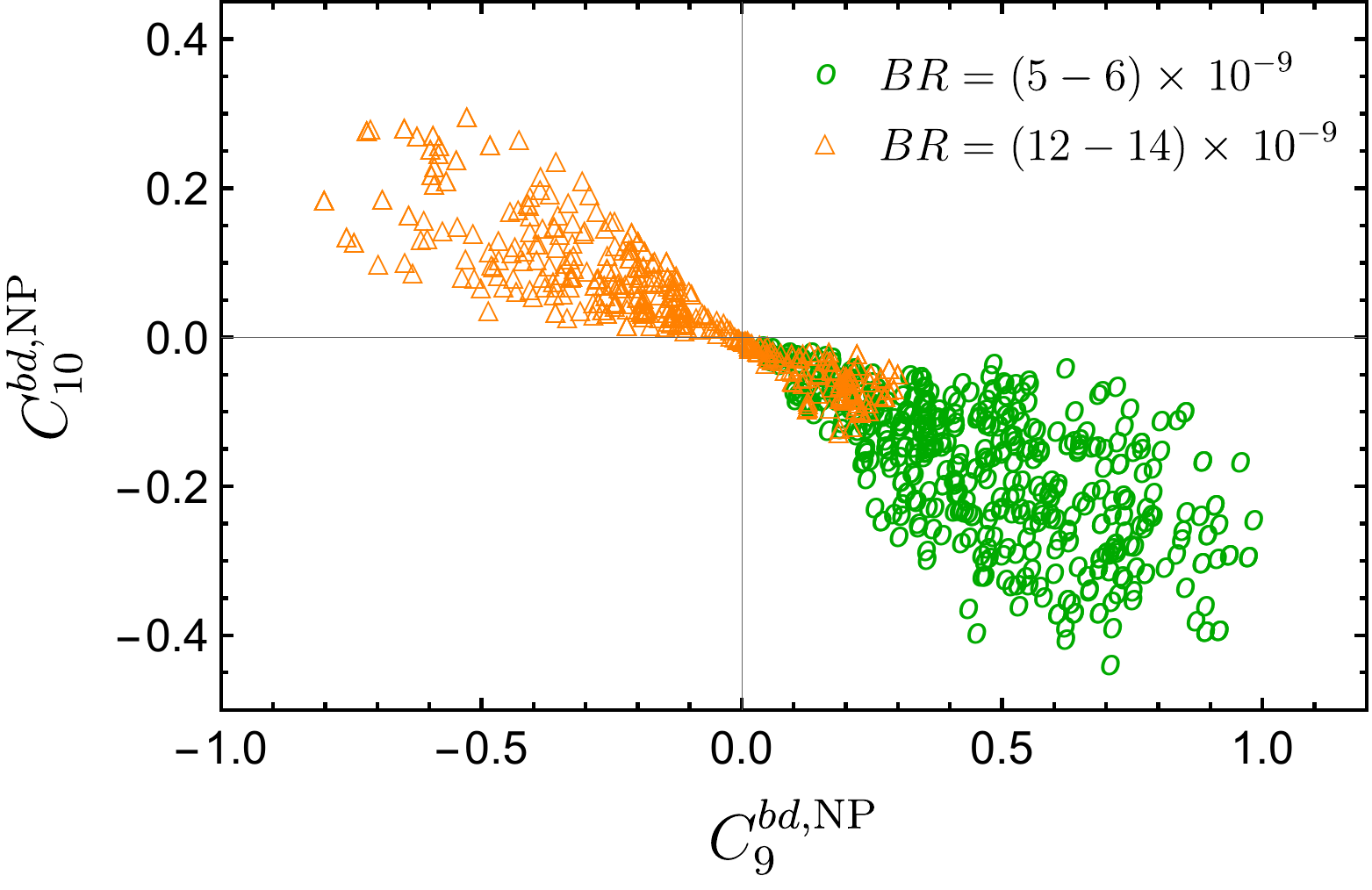}
\hspace{0.02\textwidth}
\includegraphics[width=0.46\textwidth]{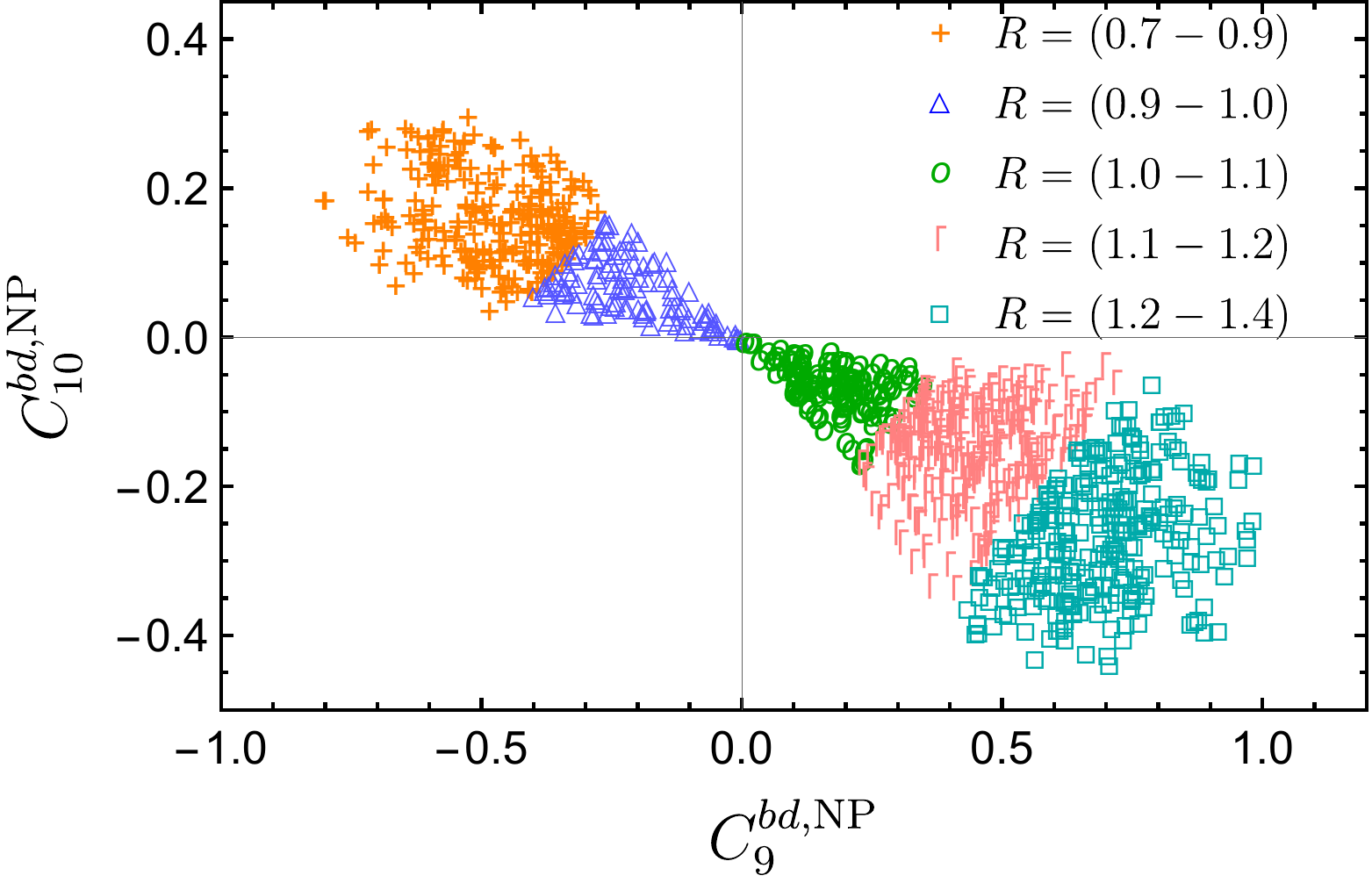}
\caption{Integrated values of the branching ratio of $B_s \to \bar{K}^* \mu^+ \mu^-$ (left panel) and the LFUV ratio \Rks (right panel)
  in the $1\sigma$-favored parameter space of $(C_9^{bd, \rm NP},\,C_{10}^{bd, \rm NP})$.}
\label{fig:re-rI}
\end{figure*}

\subsection{Real couplings}
In order to quantify how well the $Z'$ model is able to account for all data in the $b \to s$ and $b \to d$ sectors, we define $\Delta \chi^2 = \chi^2_{\rm SM} - \chi^2_{\rm NP}$,
where the minimum $\chi^2$ in the SM, and in the presence of NP $Z'$ couplings, is denoted by $\chi^2_{\rm SM}$ and $\chi^2_{\rm NP}$, respectively. For the case of real couplings,
we find the best fit values to be $g_L^{bd} =\pm\, 0.3\times 10^{-3} $, $g_L^{\mu \mu} =\mp\, 0.4  $, and $g_R^{\mu \mu} = \mp\, 0.2$.    
The value of $\chi^2_{\rm SM} \approx 221$ and $\Delta \chi^2 \approx 41$. 
The value of  $\chi^2_{\rm SM} = 221$ corresponds to $g_L^{bs} =g_L^{bd} = g_L^{\mu \mu} = g_R^{\mu \mu} =0$. Since allowing all these NP couplings to be non-zero can decrease the $\chi^2$ to $\chi^2_{{\rm NP},(s,d)}  \approx 181$, the SM point may be said to be  highly disfavoured. However, even if we restrict $g_L^{bd} =0$, the freedom allowed in the other NP couplings can still allow $\chi^2_{{\rm NP},s} \approx 181$. Thus, the improvement over the SM, $\Delta \chi^2 \approx 41$ is mainly due to the presence of non-zero  $g_L^{bs}$ and the muon couplings, which help explain the anomalies in the $b\to s \ell \ell$ sector.

The 1$\sigma$-favored parameter space of the couplings ($\gbd$, $g_{L}^{\mu \mu}$, $g_R^{\mu \mu}$) is shown in Fig.~\ref{parameterS-real}. It can be seen from
($\gbd$, $g_L^{\mu\mu}$) and ($\gbd$, $g_R^{\mu\mu}$) planes that, while $g_R^{\mu\mu} = 0$ is barely disfavored within 1$\sigma$, a rather wide strip $|g_L^{\mu \mu}| \leq 0.25$
lies beyond the 1$\sigma$-favored region. 
This is because the anomalies in $b \to s \mu \mu$ decays need a non-zero value of $C_9^{bs, \rm NP}$, which in turn require a non-zero value of $g_L^{\mu \mu}$
or $g_R^{\mu \mu}$. Furthermore, the scenario $C_9^{bs,\rm NP}=-C_{10}^{bs,\rm NP}$ \cite{Alok:2019ufo}, which provides a good fit, favors $g_R^{\mu\mu} = 0$, thus requiring $g_L^{\mu \mu}$ to be 
away from zero. 
Note that the results in the ($g_L^{\mu\mu}$ , $g_R^{\mu \mu})$ plane indicate the class of favored solutions that lie along $g_L^{\mu\mu} = g_R^{\mu \mu}$, corresponding to $C_{10}^{bs,\rm NP} \approx 0 \approx C_{10}^{bd,\rm NP} $.

\subsubsection{Predictions for $dB/dq^2$, $R_{K^*}^{(s)}(q^2)$, $A_{FB}(q^2)$ and $F_L(q^2)$}

The  top left panel of Fig.~\ref{fig:re-pred} shows predictions for the differential branching ratio corresponding to real $Z'$ couplings, for the SM as well as two benchmark scenarios NP1 and NP2
from Table.~\ref{table-imwc}. These scenarios roughly correspond to the maximum deviation on either side from the SM predictions in the $1\sigma$ favored NP parameter space.  
The maximum enhancement (suppression) in the differential branching ratio corresponds roughly to a maximum positive (negative) value of $C_9^{bd, \rm NP}$.
It can be seen from the figure that only a marginal enhancement or suppression over the SM value is possible in the differential branching ratio. 
A clean distinction among the predictions of different scenarios is difficult owing to the large uncertainties (about 20\%) arising from the form-factors.

A measurement of the LFUV ratio $R_{K^*}^{(s)}(q^2)$ in a few $q^2$ bins would be possible with the LHCb upgrade-II data set \cite{Cerri:2018ypt}. The predictions for 
this quantity in the benchmark scenarios NP1 and NP2 are shown in the top right panel of Fig.~\ref{fig:re-pred}. In the SM, $R_{K^*}^{(s)}(q^2)$ is unity in the entire low-$q^2$ region, 
while an enhancement up to 1.3 and a suppression up to 0.8 is allowed. The maximum enhancement (suppression) roughly corresponds to the maximum positive (negative)
value of $C_9^{bd,\rm NP}$.

Within the SM, the forward-backward asymmetry $A_{FB}(q^2)$ is predicted to vanish around $q^2$ $\approx$ $3.5\,\mathrm{GeV^2}$, and the zero-crossing is from negative to positive,
as can be seen from the bottom left panel in Fig.~\ref{fig:re-pred}. The maximum value of $A_{FB}(q^2)$ in the SM is $\approx$ 10 \%.
The positive (negative) value of $C_9^{bd, \rm NP}$ also shifts the zero-crossing towards lower (higher) $q^2$ value.
The integrated value of $A_{FB}$ over $q^2 =(1-6)\,\rm{GeV^2}$ bin is $(-0.6\pm 1)\,\%$ within the SM . The predictions for integrated $A_{FB}$ for the benchmark scenarios NP1 and NP2 are
$(3.1 \pm 1.4)\%$ and $(-5 \pm 1.7)\%$, respectively.

The predictions for longitudinal polarization fraction $F_L(q^2)$ are shown in the bottom right panel of Fig.~\ref{fig:re-pred}. Within the SM, the peak value of $F_L(q^2)$ is
$\approx$ 0.9 around $q^2 \approx 1.8 $ $\rm GeV^2$. The shape of $F_L(q^2)$ does not change with NP and only a marginal deviation from SM is allowed for the benchmark NP scenarios considered here.

Thus, in the case of real couplings, $R_{K^*}^{(s)}(q^2)$ is useful to distinguish the predictions of the two benchmark NP scenarios from the SM expectation, while 
the predictions for the differential branching ratio, $A_{FB}(q^2)$, and $F_L(q^2)$ may not have distinct NP signatures,
owing to the large form factor uncertainties.

\begin{figure*}[h]
\includegraphics[width=0.4\textwidth]{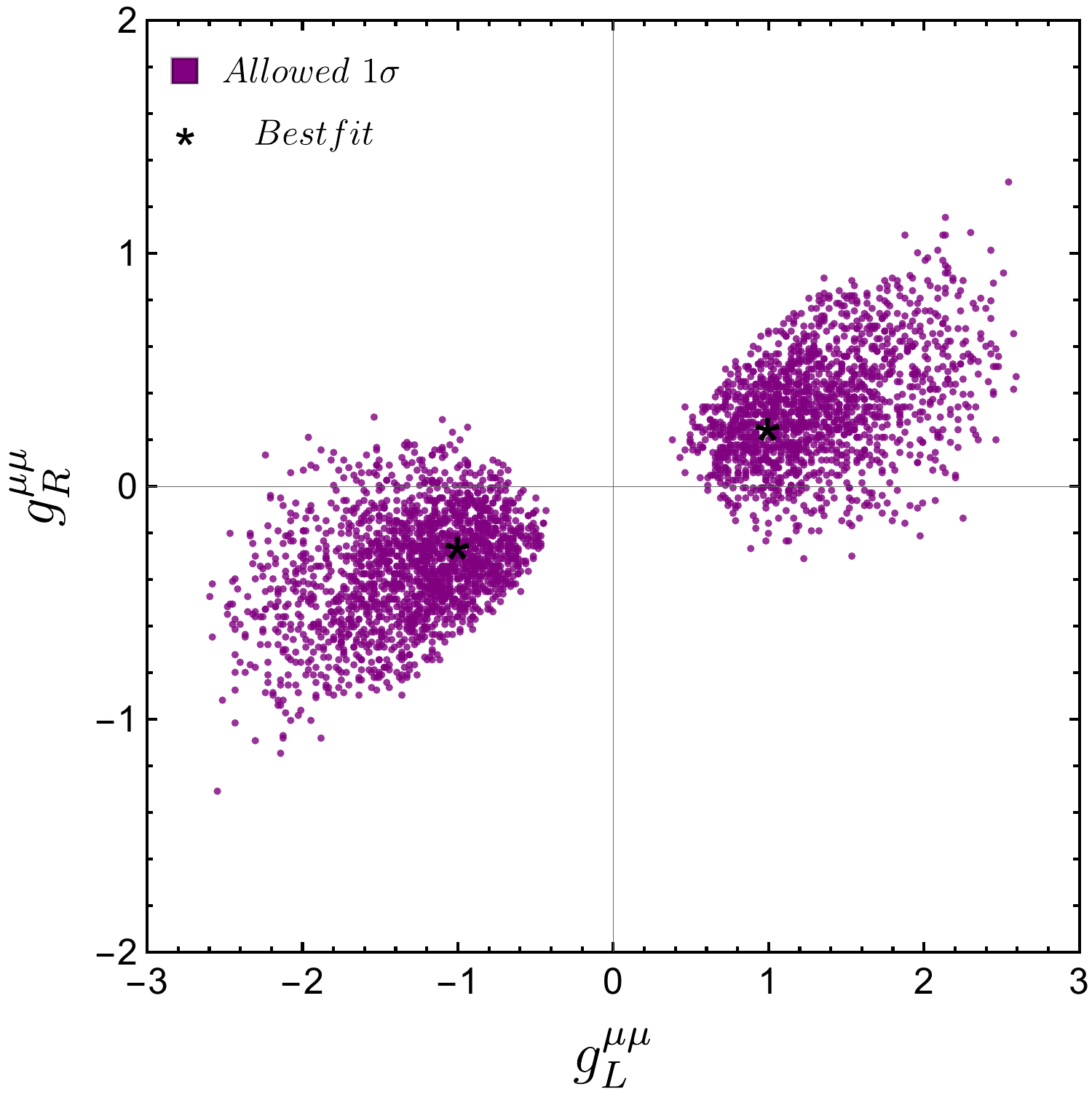}
\hspace{0.6cm}
\includegraphics[width=0.4\textwidth]{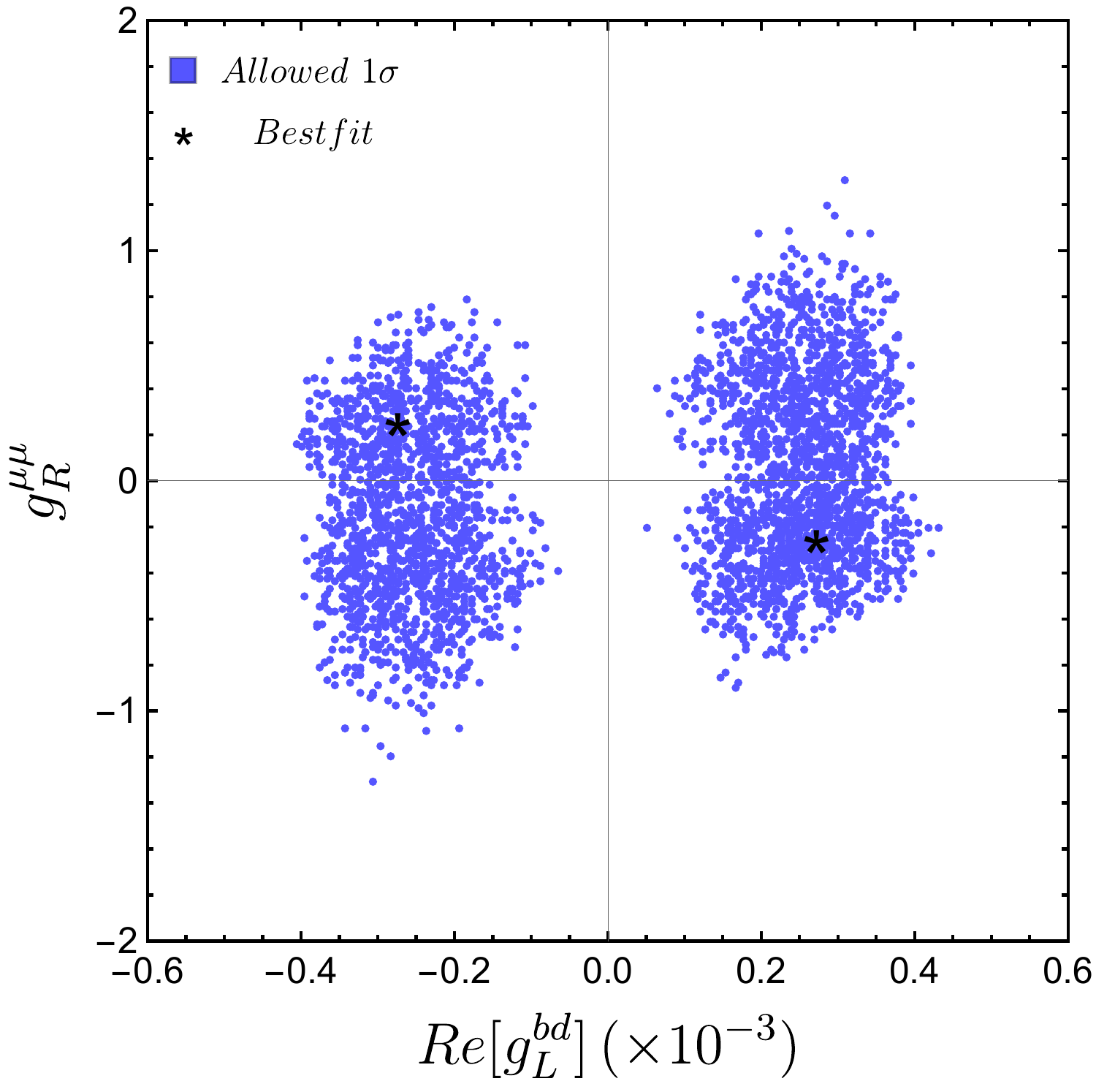}\\
\vspace{0.4cm}
\includegraphics[width = 0.4\textwidth]{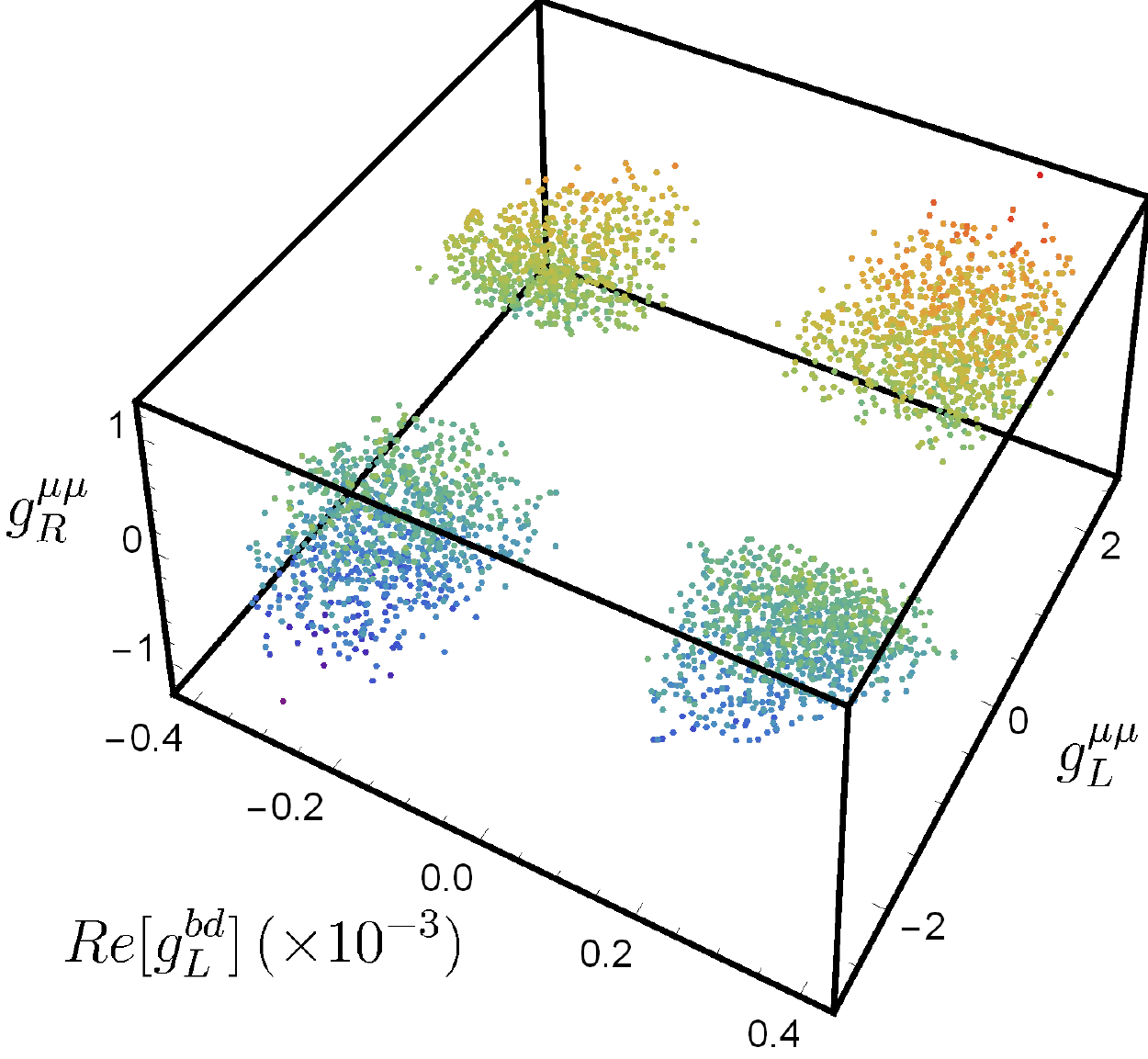}
\hspace{0.6cm}
\includegraphics[width=0.4\textwidth]{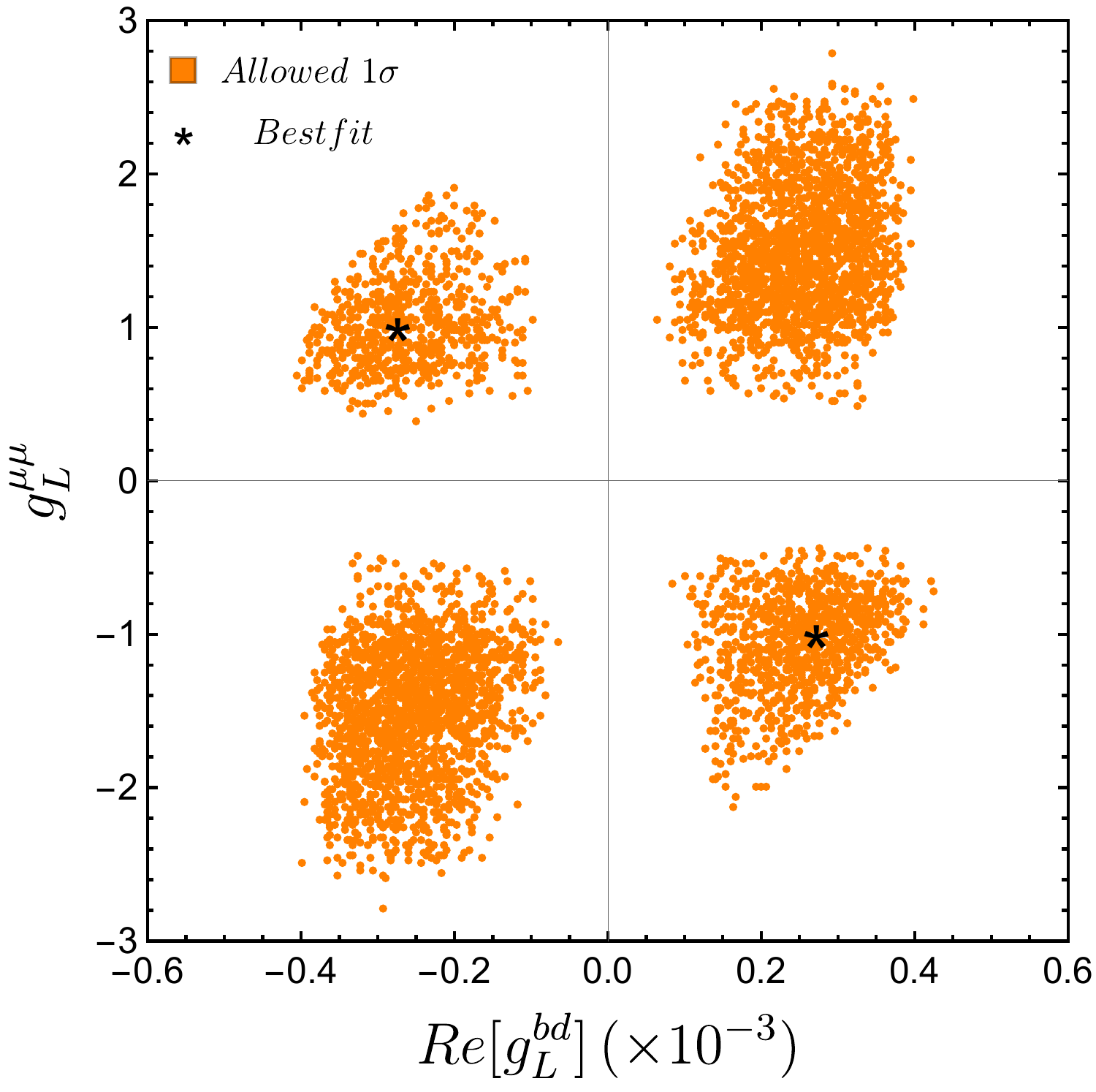}
\caption{The 1$\sigma$-favored (Re[$g^{bd}_L] $, $g_{L}^{\mu\mu}$, $g_{R}^{\mu\mu}$) parameter space for a $Z'$ model with complex couplings, for $M_{Z'}=1$ TeV. The colors red to blue in the bottom
left 3D parameter space correspond to decreasing values of $g_R^{\mu \mu}$. }
\label{fig:complexgbdglgr}
\end{figure*}

\subsubsection{Integrated branching ratio and \Rks in the low-$q^2$ region}
The results obtained for integrated $dB/dq^2$  and $R_{K^*}^{(s)}(q^2)$ over the $q^2 = (1-6)\,\mathrm{GeV}^2$ bin are presented in Fig.~\ref{fig:re-rI}. 
These results are depicted in the ($C_9^{bd,\rm NP}, C_{10}^{bd,\rm NP})$ plane, with different colors and symbols indicating
the values of integrated branching ratio (left panel) and integrated \Rks (right panel). At each 1$\sigma$-favored value of ($C_9^{bd,\rm NP}, C_{10}^{bd,\rm NP})$, we vary the values of form factor parameters
within their 1$\sigma$ range~\cite{Straub:2015ica} with a gaussian distribution of uncertainties.

In the case of integrated branching ratio, the errors due to form factors are about $20\%$. 
Due to such large errors, even by considering branching ratio values as different as (5--6)$\, \times 10^{-9}$ and (12--14)\,$\times 10^{-9}$, we find a significant overlap in the $(C_9^{bd,\rm NP}, C_{10}^{bd,\rm NP})$
plane. Hence, a measurement of integrated branching ratio may not be very helpful to put limits on the
allowed values of the NP couplings. 

In the case of $R_{K^*}^{(s)}$, the uncertainties due to form factors cancel in the ratio. The lack of overlap between the regions of integrated \Rks values in the range (0.7--1.4) indicates that a 
future measurement of integrated \Rks with an accuracy of $\sim 10\%$ in this decay mode would
make it possible to identify the ranges of $(C_9^{bd,\rm NP}\,,\, C_{10}^{bd,\rm NP})$ more precisely. Even with a preliminary measurement, an enhancement in the value of \Rks above unity
would indicate a positive
value of $C_9^{bd,\rm NP}$ and a negative value of $C_{10}^{bd,\rm NP}$, while a suppression would imply a negative
$C_9^{bd,\rm NP}$ and positive $C_{10}^{bd,\rm NP}$. This feature may be understood from the approximate analytic form of the LFUV ratio, $R_{K^*}^{(s)} \propto ({\rm Re}[C_9^{bq, \rm NP}] - {\rm Re}[C_{10}^{bq, \rm NP}])$ \cite{Hiller:2014ula}.

\begin{figure*}[t]
\includegraphics[width=0.45\textwidth]{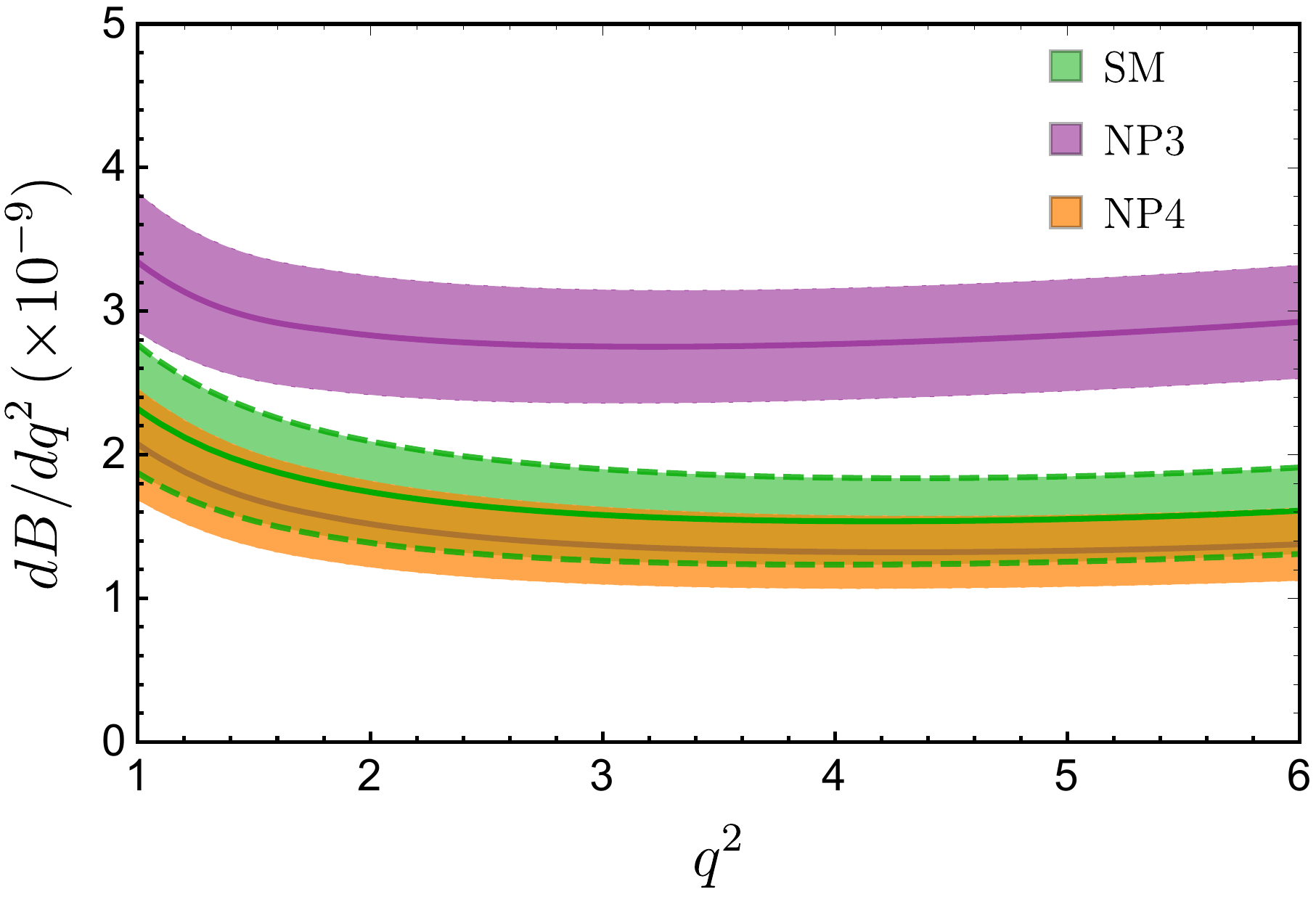}
\hspace{0.05\textwidth}
\includegraphics[width=0.455\textwidth]{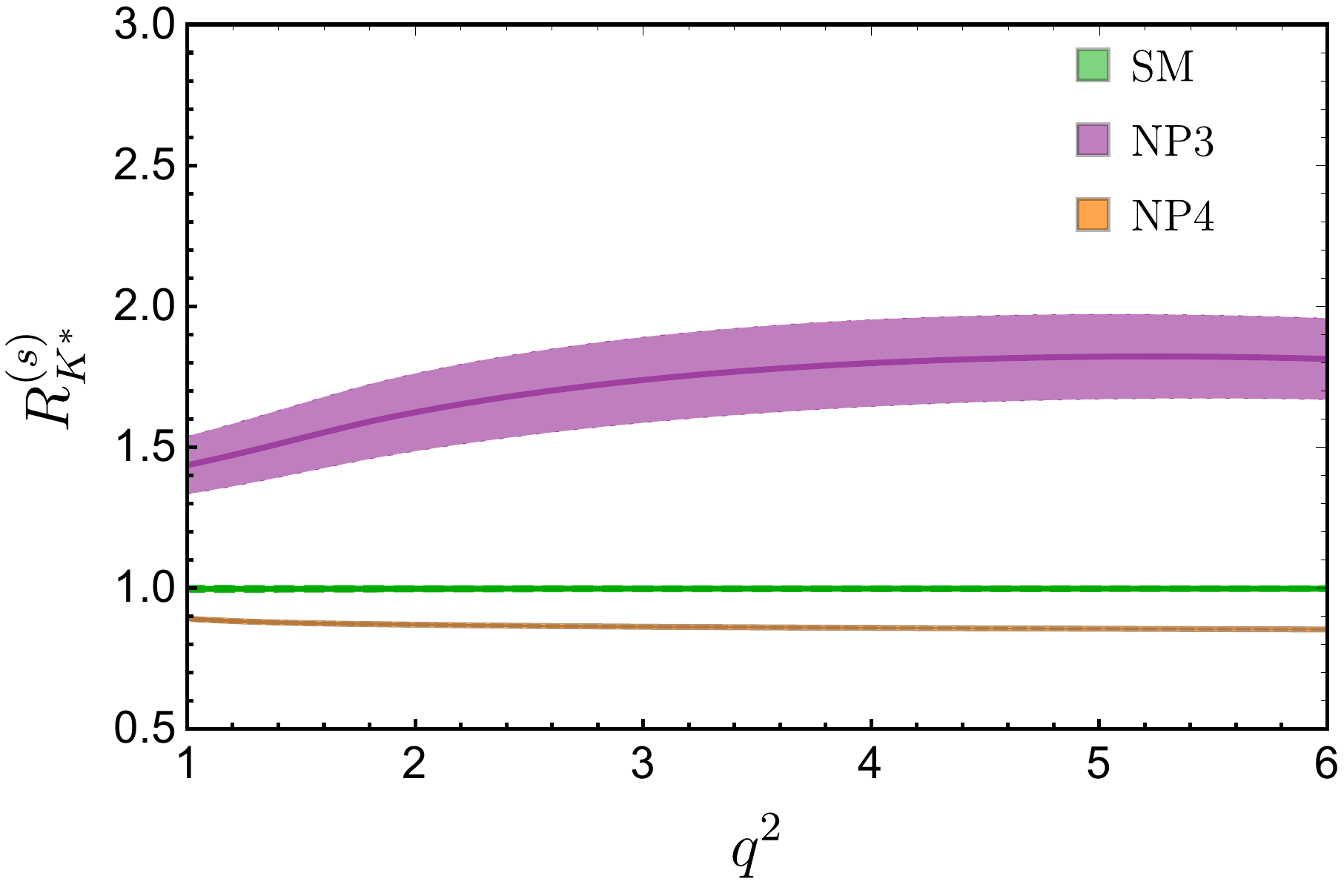}
\\ 
\includegraphics[width=0.46\textwidth]{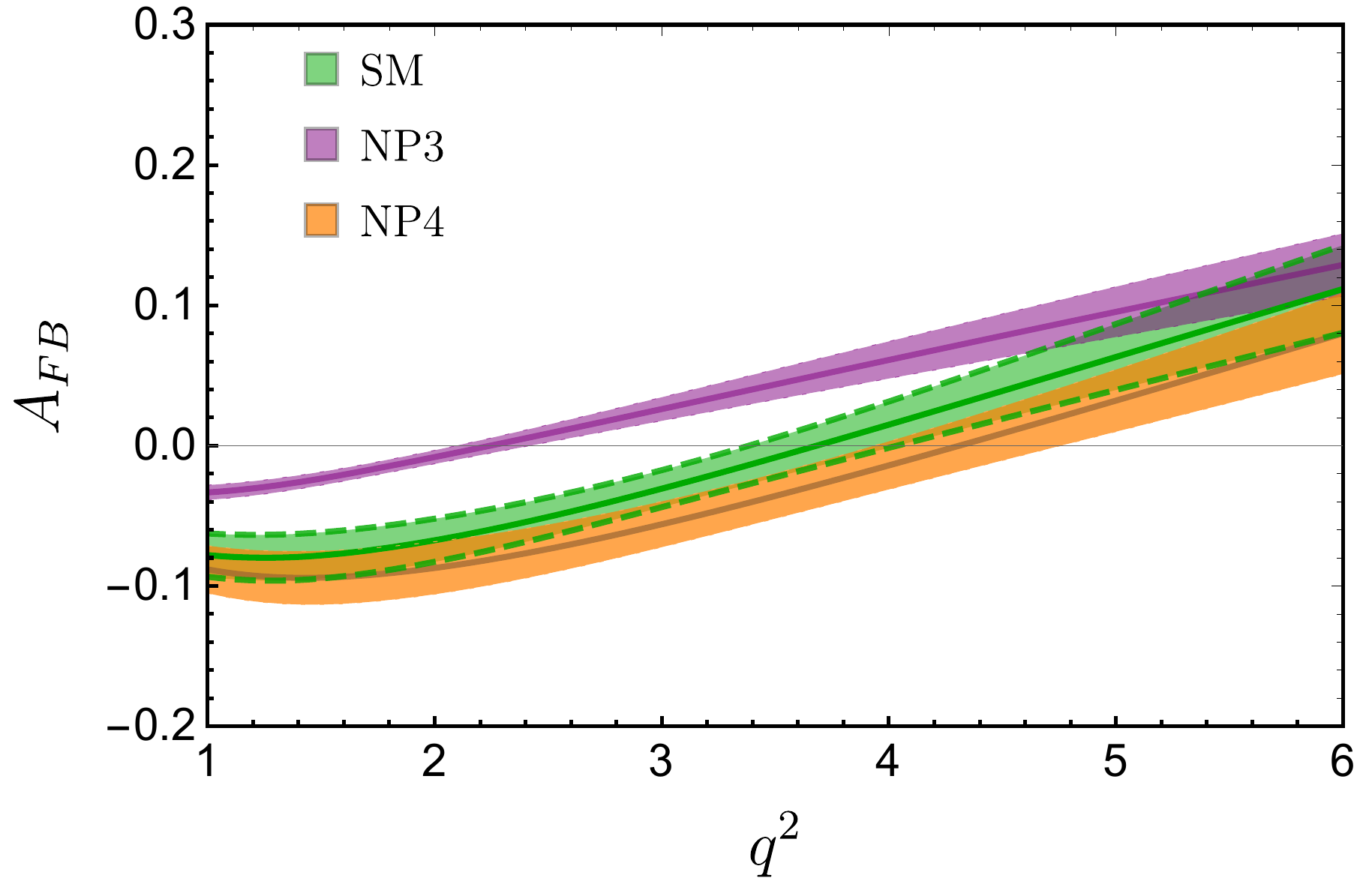}
\hspace{0.05\textwidth}
\includegraphics[width=0.45\textwidth]{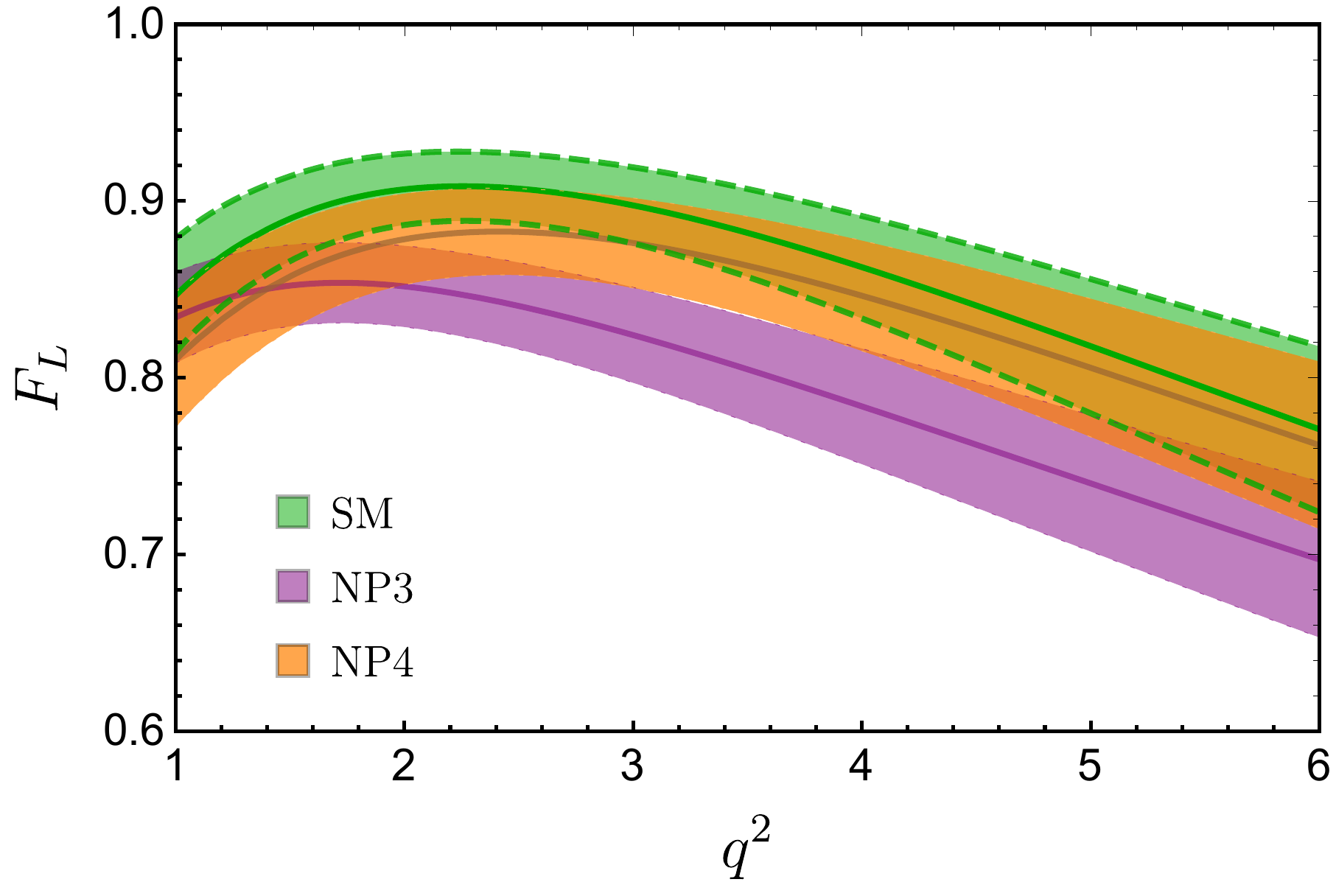}
\caption{The predictions for $B_s \to \bar{K}^*\mu^+ \mu^-$ decay, with complex $Z'$ couplings, for SM and the benchmark scenarios NP3 and NP4 as given in Table~\ref{table-imwc}\,.}
\label{fig:im-pred}
 \end{figure*}

\subsection{Complex couplings}

We would now like to see how the predictions for the above observables in the $B_s \to \bar{K}^* \mu^+ \mu^- $ decay would change if the couplings $\gbd$ and $\gbs$ are allowed to be complex.
Note that since the leptonic current in eq.~(\ref{Leff}) is self-conjugate, $g_L^{\mu \mu}$ and $g_R^{\mu \mu}$ must be real. We also study the impact of these complex couplings on the direct $CP$ asymmetry in this decay.

Fig.~\ref{fig:complexgbdglgr} shows the 1$\sigma$-favored regions of the couplings Re$[\gbd]$, $g_L^{\mu\mu}$, and $g_R^{\mu \mu}$ .
The minimum $\chi^2$ in the presence of the complex NP couplings is $ \chi^2_{\rm NP} \approx 178$, so that $\Delta \chi^2 \approx 43$, thereby providing a slightly better fit as
compared to the case of real couplings ($\Delta \chi^2 = 41$). The corresponding best fit values are Re$[\gbd] =\pm\,2.7\times 10^{-3} $,
Im$[\gbd]=\mp\,3.8\times 10^{-3} $, $g_L^{\mu \mu}=\mp\,1 $ and $g_R^{\mu\mu} =\mp\, 0.255 $. As $\chi^2_{\rm NP}$ for complex couplings is lower compared to that 
for real couplings, the 1$\sigma$-favored parameter space shifts further away from the SM point. 
A larger parameter space
is allowed for the muon couplings compared
to the real case, $|g_L^{\mu \mu}| \leq 2.5$ and $|g_R^{\mu \mu}| \leq 1.4 $. The $1\sigma$ favored region encompasses $g_R^{\mu\mu} = 0$,  whereas a rather large region around
$g_L^{\mu\mu} = 0 $ (i.e. $|g_L^{\mu\mu}| \leq 0.4$) is disfavoured within $1\sigma$. The allowed range of Im$[g_L^{bd}]$ is qualitatively similar to that of Re$[g_L^{bd}]$. 
Note that the complex nature of $g_L^{bd}$ is constrained only from $B_d- \bar{B}_d$ mixing measurements, since no $CP$-violating measurements are currently available in the $b\to d \mu \mu$ sector.

\begin{figure*}[t]
\includegraphics[width=0.46\textwidth]{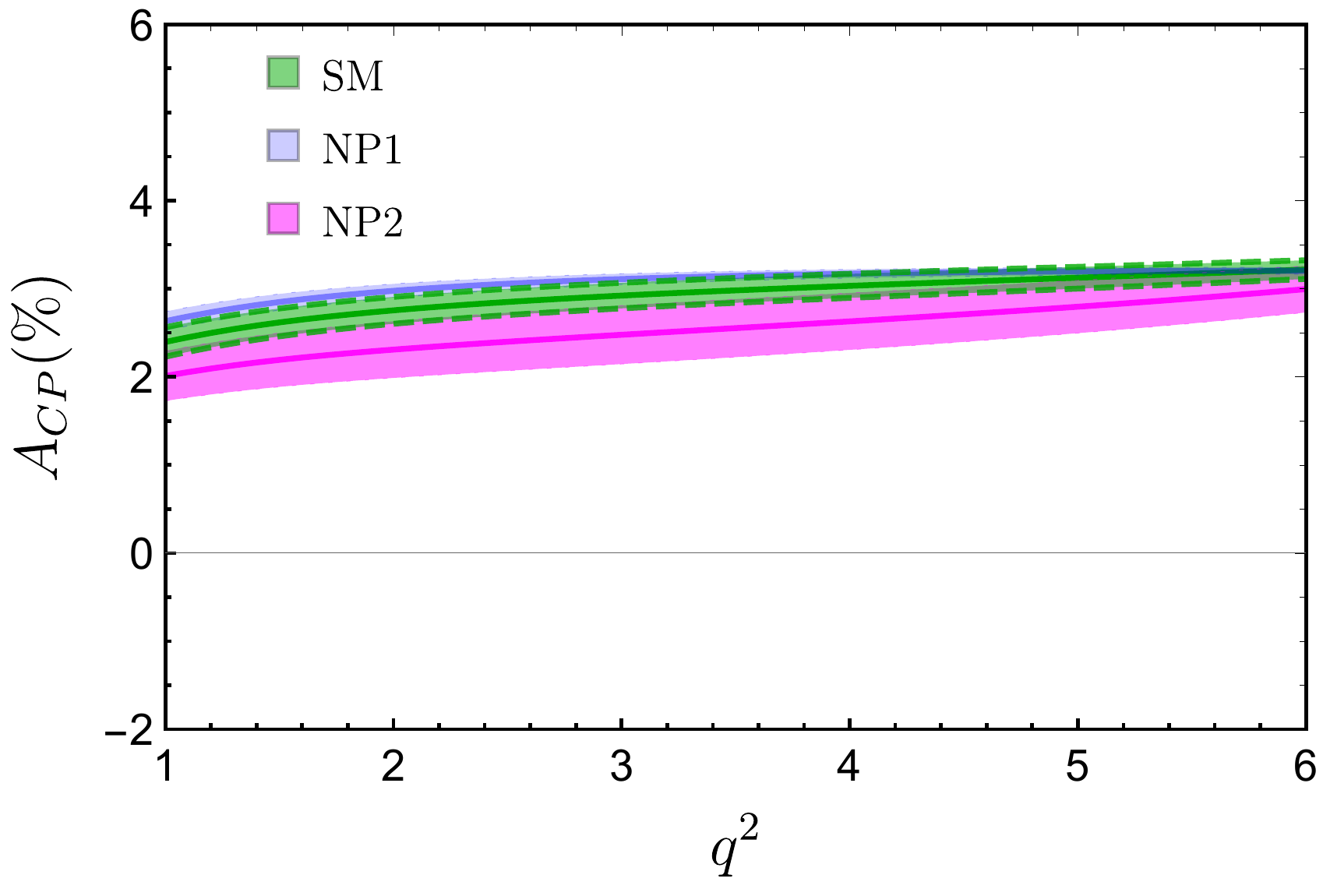}
\hspace{0.02\textwidth}
\includegraphics[width=0.46\textwidth]{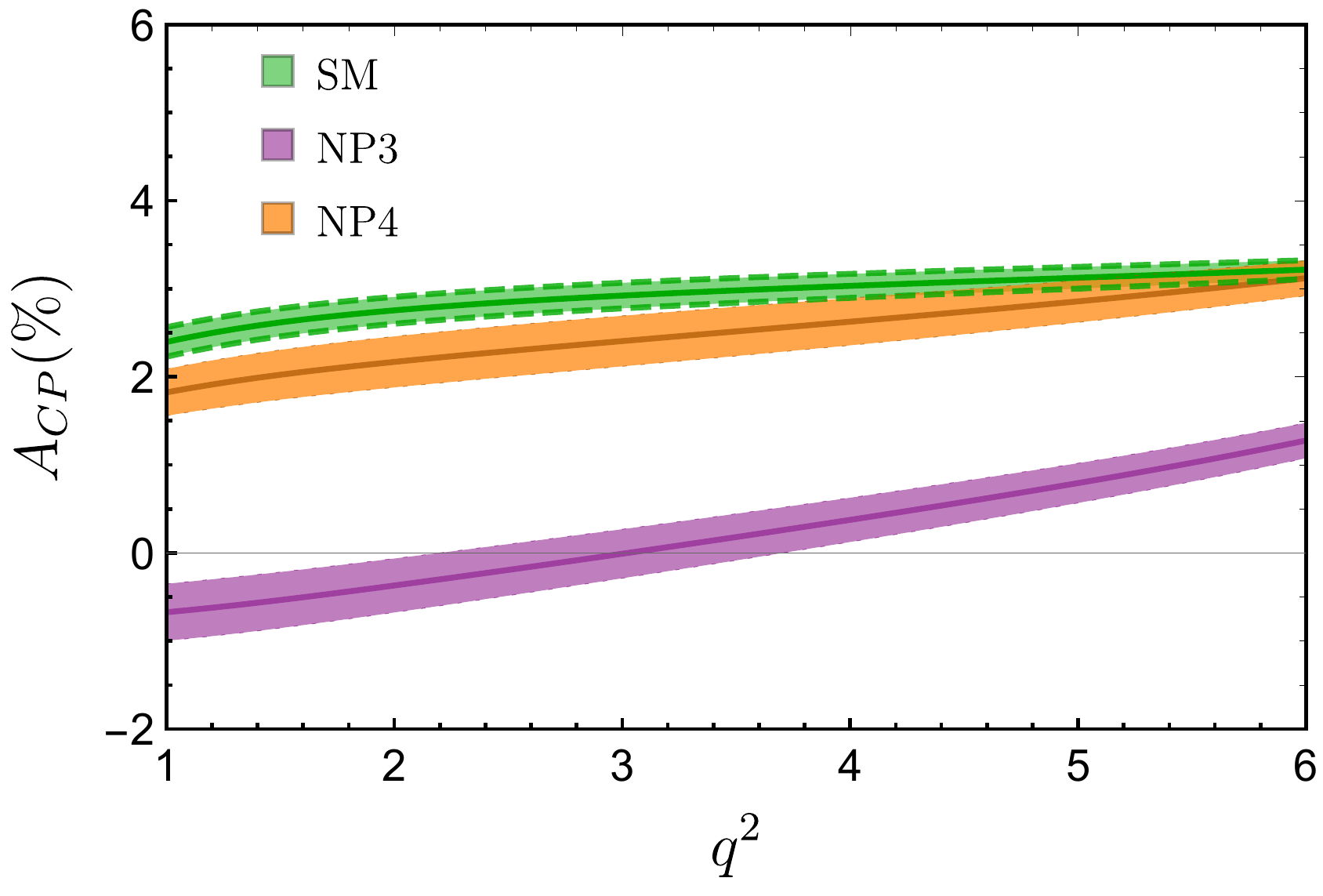}
\caption{Predictions for the direct $CP$ asymmetry $A_{CP}(q^2)$ in $B_s \to \bar{K}^* \mu^+ \mu^-$ decay for benchmark scenarios with real couplings (left)
  and complex couplings (right).} 
\label{fig:im-acp}
\end{figure*}

\begin{figure*}
\includegraphics[width=0.46\textwidth]{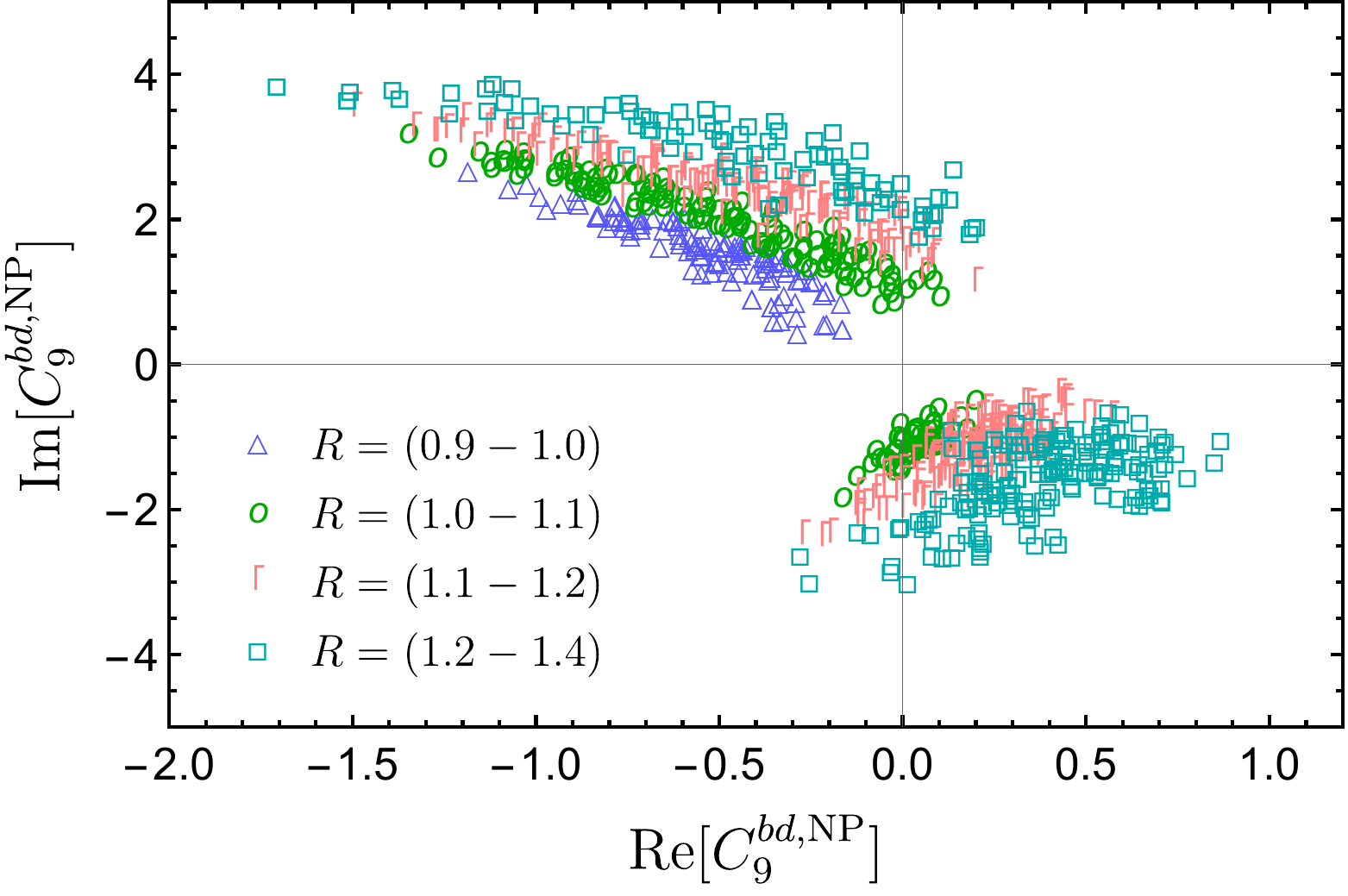}
\hspace{0.02\textwidth}
\includegraphics[width=0.46\textwidth]{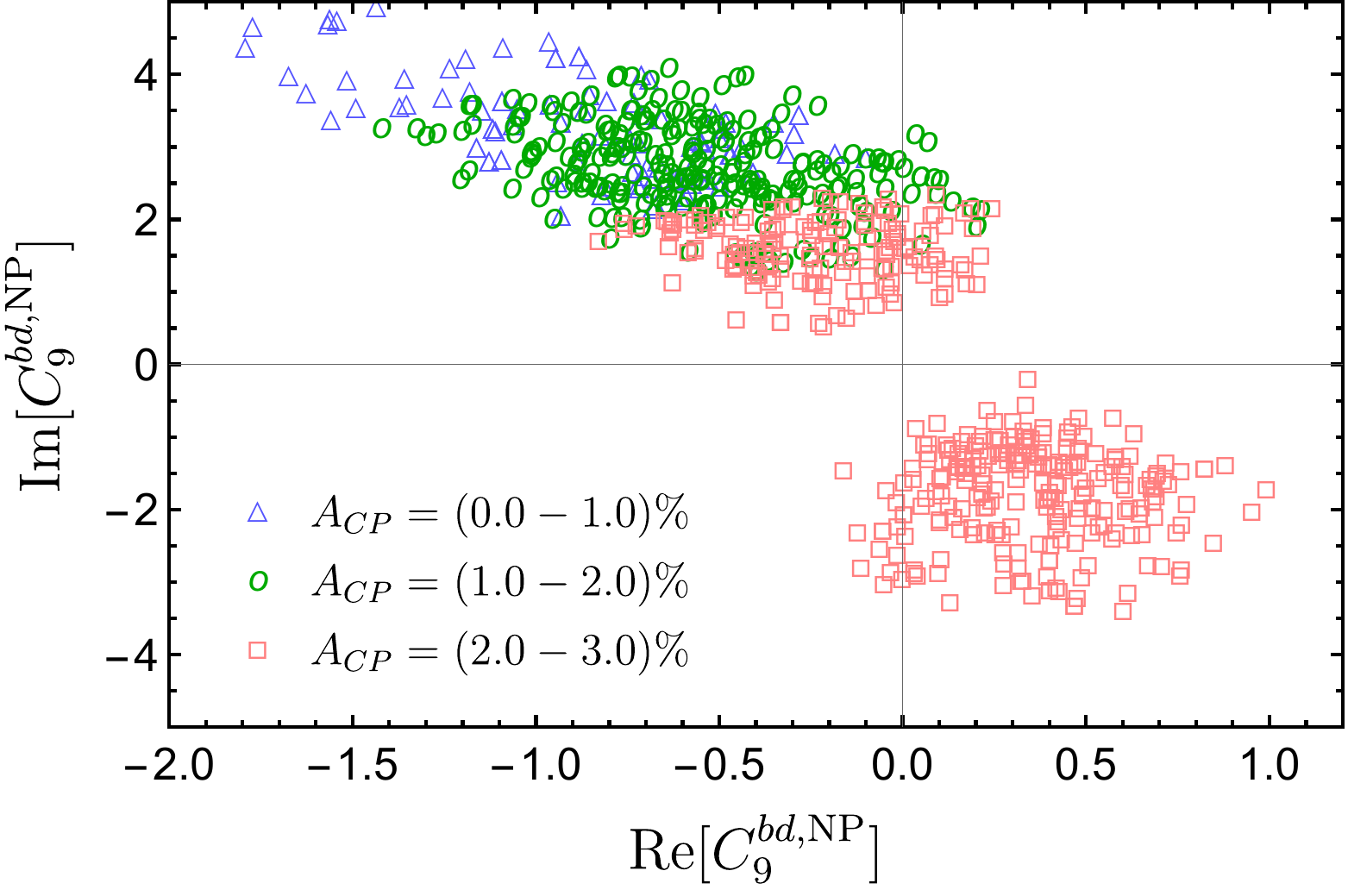}
\caption{The integrated values of the LFUV ratio \Rks (left panel) and the direct $CP$ asymmetry $A_{CP}$ (right panel)
  in the 1$\sigma$-favored parameter space of $({\rm Re}[C_9^{bd,\rm NP}],\,{\rm Im}[C_{9}^{bd,\rm NP}])$ for the $Z'$ model with complex couplings.}
\label{fig:im-racp}
 \end{figure*}
 
\subsubsection{Predictions for $dB/dq^2$, $R_{K^*}^{(s)}(q^2)$, $A_{FB}(q^2)$ and $F_L(q^2)$}
The predictions for differential branching ratio and $R_{K^*}^{(s)}(q^2)$ for the $Z'$ model with complex couplings are shown in the top panel of Fig.~\ref{fig:im-pred}\,, for SM as well as
the benchmark scenarios NP3 and NP4 in Table~\ref{table-imwc}.
These scenarios are the 1$\sigma$-favored ones with a maximum value of Im$[C_9^{bd,\rm NP}]$ and a minimum value of Re$[C_9^{bd,\rm NP}]$, respectively, and are observed to provide close to maximal allowed deviation
from the SM predictions. A significant enhancement in the branching ratio is possible in NP3, which could be useful in identifying deviations from the SM. 
A large enhancement is also possible in the LFUV ratio $R_{K^*}^{(s)}(q^2)$ in the NP3 scenario, with the maximum value of $R_{K^*}^{(s)} = 1.8$ at $q^2 = 6\,\mathrm{GeV}^2$. 
While the scenario NP4 cannot be distinguished from the SM using only the branching ratio, the value of $R_{K^*}^{(s)}(q^2)$ in this scenario can be as low as 0.85. Therefore,  $R_{K^*}^{(s)}(q^2)$ would be
useful to identify deviations from the SM.

A marginal enhancement in $A_{FB}(q^2)$ is possible for the scenario NP3, which would also display zero-crossing at much lower $q^2$ values ($q^2 \approx 2.5\,\mathrm{GeV^2}$)
compared to that in the SM ($q^2 \approx 3.5\,\mathrm{GeV^2}$). A marginal suppression in $F_L(q^2)$ is also possible in NP3.
The scenario NP4, on the other hand, does not show significant deviations from the SM for these two observables.

\subsubsection{Direct $CP$ asymmetry $A_{CP}(q^2)$}

The direct $CP$ asymmetry in the $b\rightarrow d\, \mu^+\,\mu^-$ sector is expected to be about an order of magnitude larger than $b\rightarrow s\, \mu^+\,\mu^-$.
As direct $CP$ violation in $b\rightarrow s\, \mu^+\,\mu^-$ sector is  expected to be $\sim 0.1\%$, its experimental observation would be possible only if some new physics provides an order of magnitude
enhancement to bring it up to the level of a few percent. In $b\rightarrow d\, \mu^+\,\mu^-$ decays,  the $A_{CP}$ in SM itself is at the level of a few per cent, and can be within experimental reach.

Fig.~\ref{fig:im-acp} shows $A_{CP}(q^2)$ in the low-$q^2$ region for the decay $B_s \to \bar{K}^* \mu^+ \mu^-$, considering the benchmark scenarios NP1 and NP2 (real couplings), as well as
NP3 and NP4 (complex couplings). It can be seen from the left panel of the figure that for real couplings,  $A_{CP}(q^2)$ is either marginally below the SM prediction
or almost consistent with it.

For complex couplings, however the suppression in $A_{CP} (q^2)$ can be quite large. It can even lead to $A_{CP}(q^2)$ falling below a per cent level, hence making its measurement extremely difficult.
In some scenarios (e.g. NP3), it is even possible for $A_{CP}(q^2)$ to be negative for very low $q^2$ values. After scanning over the 1$\sigma$-favored parameter space, we find no significant enhancement
in $A_{CP}(q^2)$. So an NP signal can be established if the measurements put an upper bound which is firmly below the SM prediction of $A_{CP}(q^2)$.

\subsubsection{Integrated \Rks and $A_{CP}$}

As observed in the case of real couplings, the integrated branching ratio does not help much in narrowing down the range of effective NP Wilson coefficients. Hence, in this section, we 
focus on the integrated values of \Rks and $A_{CP}$ over $q^2 = (1-6)\,\mathrm{GeV}^2$ bin. 
Fig.~\ref{fig:im-racp} depicts these results in the $(\rm {Re}[C_9^{bd,\rm NP}]$, $\rm {Im}[C_9^{bd,\rm NP}])$ plane, with different colors and symbols indicating
the values of integrated $R_{K^*}^{(s)}$ (left panel) and $A_{CP}$ (right panel). At each 1-$\sigma$-favored complex value of ($C_9^{bd,\rm NP}, C_{10}^{bd,\rm NP})$, we vary the values of form factor parameters
within their 1-$\sigma$ range~\cite{Straub:2015ica} with a gaussian distribution of uncertainties. 

As in the case of real NP couplings, integrated \Rks below the SM prediction of unity could indicate a negative value of Re$[C_9^{bd,\rm NP}]$. An enhancement in integrated \Rks upto (1.2 - 1.6) is possible
for 
large positive or negative values of Im$[C_9^{bd, \rm NP}]$. These features may be understood from the observation that in the case of complex couplings,
$R_{K^*}^{(s)}$ has contributions both from Re$[C_9^{bd,\rm NP}]$ and $|C_9^{bd, \rm NP}|^2$.

The right panel of Fig.~\ref{fig:im-racp} shows that a large positive value of Im$[C_9^{bd,\rm NP}]$ can decrease the integrated $A_{CP}$ to less than a per cent.  
The negative values of Im$[C_9^{bd,\rm NP}]$ do not seem to affect $A_{CP}$ much, keeping it close to the SM prediction of 2.5\%. 
Therefore, a simultaneous measurement of integrated \Rks and $A_{CP}$, with a precision of 0.1 and $1\%$, respectively, may help identify the sign of Im$[C_9^{bd, \rm NP}]$.  
We find that the measurements of integrated \Rks and $A_{CP}$ values are not very useful in identifying the allowed ranges of Re[$C_{10}^{bd,\rm NP}]$ and Im[$C_{10}^{bd,\rm NP}]$.

\section{Summary and Conclusions}
\label{sec:V}

In non-universal $Z'$ models, instrumental in accounting for the flavor anomalies, the observables in $b \to s \ell \ell $ and $b \to d \ell \ell$ processes would be correlated.  In this paper, 
we study the constraints on the couplings of a non-universal $Z'$ model from the measurements in $b \to q \mu \mu$ $(q = s,d)$ decays, $B_q-\bar{B}_q$ mixing,
and neutrino trident production. 
These couplings give rise to new additional contributions
to the Wilson coefficients $C_9^{bq}$ and $C_{10}^{bq}$. Using the above constraints, we perform a global fit to determine 1$\sigma$-favored regions in the parameter space
 of the $Z'$ couplings $g_L^{bd}$, $g_L^{bs}$, $g_L^{\mu \mu}$, and $g_R^{\mu \mu}$. We analyze the cases when quark-$Z'$ couplings $g_L^{bd}$ and $g_L^{bs}$ are 
(i) real, and (ii) complex. We also present our predictions for 
some important observables in $B_s \to \bar{K}^* \mu \mu$ decays ---
the differential branching ratio $dB/dq^2$, the LFUV ratio \Rks, the angular
observables $A_{FB}$ and $F_L$, and the $CP$ asymmetry $A_{CP}$ --- for some benchmark scenarios. 

It is observed from our analyses that the $Z'$ model improves the global fit over the SM by $\Delta \chi^2 \approx 41$ (real couplings) and $\Delta \chi^2 \approx 43$ (complex couplings). 
The favored regions in the parameter space lie along $g_L^{\mu \mu} \approx g_R^{\mu \mu}$, 
corresponding to $C_{10}^{bd, \rm NP} \approx 0$, while the region
around $g_L^{\mu \mu} = 0$ is disfavored. These are mainly dictated by the $R_K$ and $R_{K^*}$ anomalies in $b \to s$ sector.

For the observables in $B_s \to \bar{K}^* \mu^+ \mu^-$ decays, when the couplings are real, we find that the enhancement and suppresion in $dB/dq^2$ cannot be 
cleanly identified due to the large uncertainties in the SM prediction. 
However, the value of $R_{K^*}^{(s)}(q^2)$ can substantially deviate from the SM prediction of unity --- it can range from 0.8 to 1.3. The enhancement (suppression) 
corresponds to positive (negative) values of $C_9^{bd, \rm NP}$.  
 A marginal enhancement 
and suppression in $A_{FB}(q^2)$ is possible compared to the SM predictions, with the zero-crossing shifting towards lower (higher) $q^2$ values
for positive (negative) values of $C_9^{bd, \rm NP}$. There is no significant deviation from SM in the predictions of $F_L(q^2)$, and the predictions of $A_{CP}(q^2)$
also stay close to the SM expectation for all the favored values of NP Wilson coefficients. 
Further, we find that a measurement of integrated \Rks in the low-$q^2$ bin with a precision of $\sim 0.1$ can help 
narrow down the ranges of $(C_9^{bd,\rm NP}, C_{10}^{bd, \rm NP})$. 

In the case of complex couplings, a larger NP parameter space is allowed, leading to larger possible deviations in the $B_s \to \bar{K}^* \mu \mu$ observables. 
In particular, a $\sim 50\%$ enhancement in $dB/dq^2$ is allowed.
Moreover, the LFUV ratio $R_{K^*}^{(s)}(q^2)$ can be enhanced up to 1.8 in scenarios with large positive and negative Im$[C_9^{bd, \rm NP}]$. 
There can also be a significant enhancement in $A_{FB}(q^2)$ for positive values of Re$[C_9^{bd, \rm NP}]$ and large Im$[C_9^{bd, \rm NP}]$,
with the zero-crossing shifting towards lower $q^2$. 
A significant suppression in $A_{CP}(q^2)$ compared to the SM
prediction of 2.5\% is possible for large positive values of Im[$C_9^{bd,\rm NP}]$, which may lead to $A_{CP}(q^2)$ falling below a per cent level. We find that a measurement 
of integrated \Rks and $A_{CP}$, with a precision 0.1 and 1\%, respectively, would be needed to narrow down the allowed ranges of (Re[$C_9^{bd,\rm NP}]$, Im$[C_{9}^{bd,\rm NP}])$.

To summarize, we study NP effects in $B_s \to \bar{K}^* \ell \ell$ decays in a generic $Z'$ model with real as well as complex couplings.
  The constraints on the $Z'$ couplings are obtained by correlating measurements in the $b \to s$ and $b \to d$ sectors, along with neutrino trident production. We find that
\begin{itemize}
\item The present data allow a large deviation (enhancement as well as suppression) in \Rks from its SM prediction. The deviation is more pronounced for complex NP couplings.  

\item The $CP$ asymmetry can be significantly suppressed as compared to the SM prediction. 
\end{itemize}

The modes $B_s \to \bar{K}^* \ell \ell$ are expected to be measured with a good accuracy in the near future. The observables
\Rks and $A_{CP}$ in $B_s \to \bar{K}^* \mu \mu$ decays can show clean signatures of the presence of NP. Hence their measurements will be crucial in the search for physics beyond the SM.


\bigskip
\noindent
    {\bf Acknowledgements}: We would like to thank Suman Kumbhakar for his contribution during the initial stages of this work. The work of DK is supported
    by the National Science Centre (Poland)
    under the research grant No. 2017/26/E/ST2/\\00470. We would like to thank the organizers of WHEPP 2019, where this work was completed.


\appendix

\section{Constraints from $b \to s \nu \bar{\nu}$}

\begin{figure}
\centering
\includegraphics[width=0.4\textwidth]{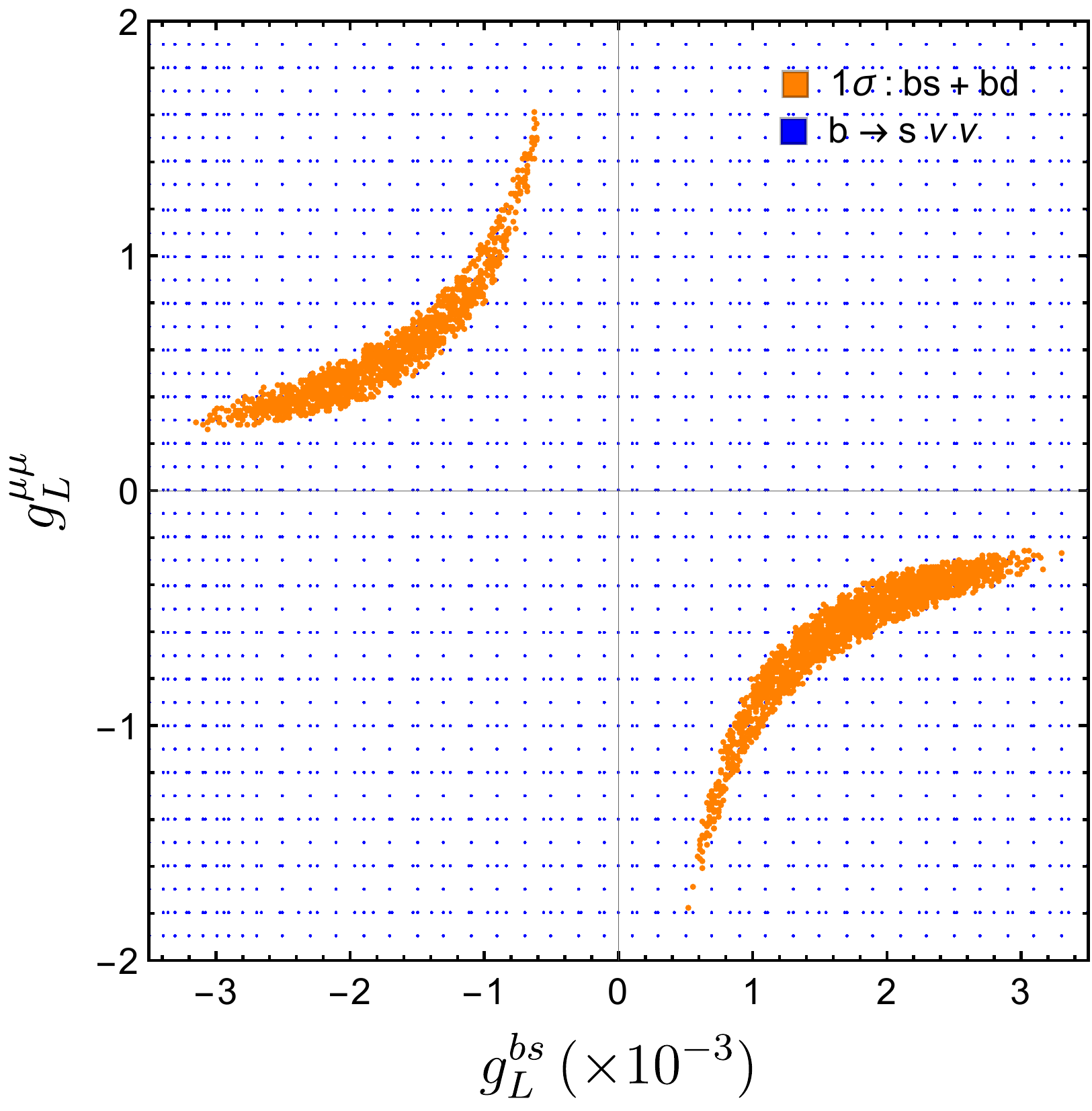}
\caption{Comparison of favored NP parameter space using  $b\to s \nu \bar{\nu}$ data and combined fit to $b\to s $, $b\to d$ and neutrino trident data.}
\label{fig:bsnunu}
\end{figure}

\begin{figure*}
\includegraphics[width=0.45\textwidth]{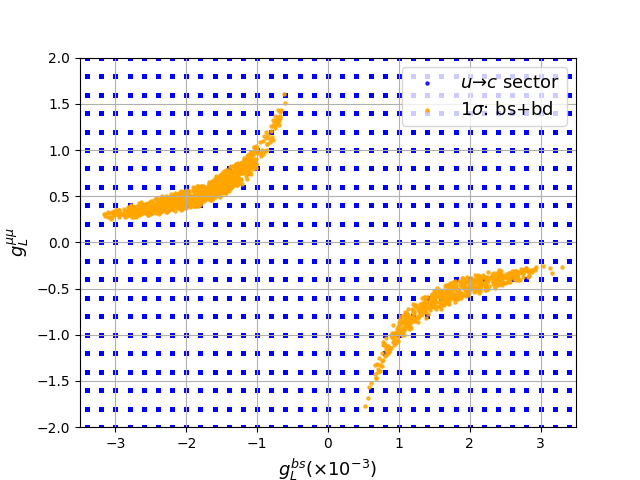}
\hspace{0.05\textwidth}
\includegraphics[width=0.45\textwidth]{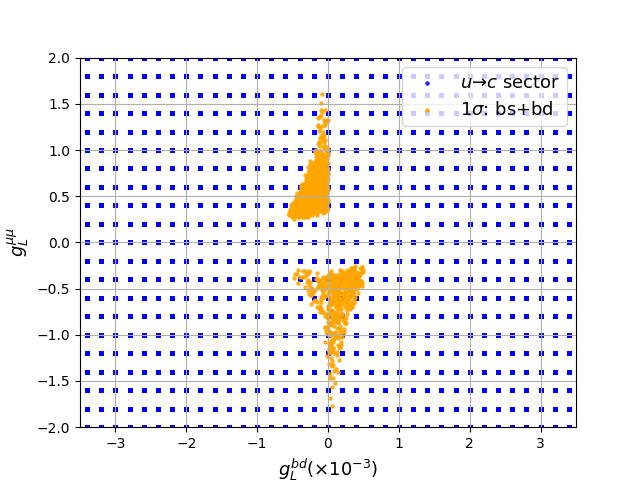}
\caption{Comparison of allowed  NP parameter space using $D^0$-$\bar{D^0}$ mixing \& branching ratio of $D^0 \to \mu^+ \mu^-$  and combined $b\to s $, $b\to d$ and neutrino trident fit.}
\label{fig:charm}
\end{figure*}

The quark level transition $b \to s \nu \bar{\nu}$ induces exclusive semi-leptonic decays
$B \to K^{(*)}\nu\nu$. The effective Hamiltonian relevant for $b\to s \nu \bar{\nu}$ transition is \cite{Buras:2014fpa}
\begin{equation}
H_{\rm eff} = - \frac{\sqrt{2} \alpha G_F}{\pi} V_{tb} V_{ts}^* \sum_\ell
C_L^\ell (\bar s \gamma_{\mu} P_L b) (\bar \nu_\ell \gamma^{\mu}
P_L\nu_\ell) ~,
\end{equation}
where $C_L^\ell = C_L^{\rm SM} + C_\nu^{\ell\ell}({\rm NP})$. The NP contribution $C_\nu^{\mu\mu}({\rm NP})$ in the $Z'$ model is  given by
  \begin{equation}
  C_\nu^{\mu\mu}({\rm NP}) =  -\frac{\pi}{\sqrt{2}G_F\alpha V_{tb}V^*_{ts}} \frac{g_L^{bs} g_L^{\mu\mu}}{M^2_{Z'}}\,.
  \end{equation}
The SM WC is $C_L^{\rm SM} = - X_t/s_W^2$, where $s_W \equiv \sin\theta_W$ and $X_t = 1.469 \pm 0.017 $.

From the experimental side, at present, we only have following upper limits \cite{Grygier:2017tzo,Lees:2013kla,Lutz:2013ftz,delAmoSanchez:2010bk,Aebischer:2018iyb}
\begin{eqnarray}
{\cal B}(B^0 \rightarrow K^0 \nu \bar \nu ) &<& 2.9 \times 10^{-5} ~, \nonumber\\
{\cal B}(B^0 \rightarrow K^{*0} \nu \bar \nu ) &<& 2.0 \times 10^{-5} ~,\nonumber\\
{\cal B}(B^+ \rightarrow K^+ \nu \bar \nu ) &<& 1.7 \times 10^{-5} ~, \nonumber\\
{\cal B}(B^+ \rightarrow K^{*+} \nu \bar \nu ) &<& 4.8 \times 10^{-5} ~.
\end{eqnarray}
Using the above bounds, the allowed NP parameter space from $b\to s \nu \bar{\nu}$ data near our best-fit region is depicted in Fig.~\ref{fig:bsnunu}. It is evident that the  bounds coming from the current $B \to K^{(*)}\nu\nu$ data are much weaker than those obtained from the combined $b\to s$, $b\to d$ and neutrino-trident fit.

\section{Constraints from  $D^0$-$\bar{D^0}$ mixing and $D^0 \to \mu^+ \mu^-$  decay }

The  quark doublets in Eq.~(2) are taken to be in the down-type quark diagonal basis. Hence owing to quark mixing, the up-type quarks in the quark doublets induce $u_i \to u_j$ transitions. Then there can be constraints coming from the up quark sector, in particular $D^0$-$\bar{D^0}$ mixing and charm decays. The
relevant terms in the effective Hamiltonian for the $c \to u$ sector are 
\begin{eqnarray}
  \mathcal{H}_{\rm eff}^{Z', {c \to u}} &\supset &  \frac{1}{2M^2_{Z'}}J_{\alpha}J^{\alpha} = \frac{g^{cu}_L}{2M^2_{Z'}}\left(\bar{u}\gamma^{\alpha}P_L c\right)\left(\bar{u}\gamma_{\alpha}P_L c\right)
  \nonumber\\
  &  +&  \frac{h^{cu}_L}{M^2_{Z'}} \left(\bar{u}\gamma^{\alpha}P_L c\right)
  \left[\bar{\mu}\gamma_{\alpha}\left(g^{\mu\mu}_L P_L 
   + g^{\mu\mu}_R P_R\right)\mu \right]\,,
 \label{Leff-charm}
\end{eqnarray}
where
\begin{eqnarray}
g^{cu}_L &=& (g_L^{bs} V_{ud} V_{cb}^{*})^2 +  (g_L^{bd} V_{us} V_{cb}^{*})^2 + 2g_L^{bs}g_L^{bd} V_{ud} V_{cb}^{*} V_{us} V_{cb}^{*}\,, \nonumber\\
h^{cu}_L &=& g_L^{bs} V_{ud} V_{cb}^{*} + g_L^{bd} V_{us} V_{cb}^{*}\,.
\end{eqnarray}
The first term in Eq.~\ref{Leff-charm} induces $D^0$-$\bar{D^0}$ mixing  whereas the second term induces $c \to u \mu^+ \mu^-$ transition. Here we consider constraints from $D$-$\bar{D}$ mixing and $D^0 \to \mu^+ \mu^-$.

In the SM, $D^0$-$\bar{D^0}$ mixing is induced at the loop 
level by the quarks d, s and b. Due to a strong GIM cancellation, 
the short-distance contribution is extremely small. In particular, the contribution 
due to  b-quark is highly suppressed, $O(\lambda^8)$. Therefore $D^0$-$\bar{D^0}$ mixing is dominated by the d- and s-quarks and hence there can be large long-distance contributions, for which there are no reliable estimates at present \cite{Petrov:2006nc,Golowich:2009ii}.  In our analysis, we consider the  $D^0$-$\bar{D^0}$ mixing parameter $\Delta M_D$ which is measured to be $0.0095^{+0.0041}_{-0.0044}$ $\rm ps^{-1}$ \cite{pdg}. In $Z'$ model, $D^0$-$\bar{D^0}$ mixing is induced at the tree level and hence would provide a much larger contribution in comparison to the short-distance SM contribution. Further as long-distance contributions are unknown, we saturate the  $\Delta M_D$ experimental value with new physics contribution which is given by
\begin{equation}
\Delta M_D = \frac{f_D^2\, m_D\, B_D\, r(m_c,M_{Z'})}{3M^2_{Z'}}(g^{cu}_L)^2 \,,
\end{equation}
where $f_D = 212.0 \pm 0.7$ MeV \cite{Aoki:2019cca}, $B_D = 0.757 \pm 0.027 \pm 0.004$ \cite{Carrasco:2015pra} and $r(m_c,M_{Z'}) = 0.72$ for $M_{Z'} = 1$ TeV \cite{Golowich:2007ka}.

The decay $D^0 \to \mu^+ \mu^-$ is induced by the quark level transition $c \to u \mu^+ \mu^-$. In $Z'$ model,  the branching ratio is given by
\begin{align}
{B}(D^0 \to \mu^+ \mu^-) &= \frac{\tau_D f_D^2 m_{\mu}^2 m_D}{32 \pi M^4_{Z'}}\sqrt{1-\frac{4m_{\mu}^2}{m_D^2}}\nonumber\\
& \times (h^{cu}_L)^2 (g_L^{\mu\mu}-g_R^{\mu\mu})^2 \,.
\end{align}
Within the SM, $D^0 \to \mu^+ \mu^-$  is dominated by the intermediate $\gamma^* \gamma^*$ state which scales its branching ratio as $2.7 \times 10^{-5}$ 
times the branching ratio for $D^0 \to \gamma \gamma$ \cite{Burdman:2001tf}. Using the upper bound on $D^0 \to \gamma \gamma < 2.2 \times 10^{-6}$   at 90\% C.L. \cite{Lees:2011qz}, the SM branching ratio  is estimated to be $\lesssim 10^{-10}$. From the experimental side, we only have an upper bound which is $  < 6.2 \times 10^{-9}$ at 90\% C.L. \cite{Aaij:2013cza}.

Fig.~\ref{fig:charm} shows the region allowed by the branching ratio of $D^0 \to \mu^+ \mu^-$ and $\Delta M_D$  as well as  from the combined fit in the region around our best-fit point. It is obvious that the constraints coming from the charm sector are significantly weaker.



\end{document}